\nonstopmode

\newcommand{\ie}{i.e.{}}
\newcommand{\eg}{e.g.{}}

\newcommand{\eV}{\U{eV}}

\newcommand{\mum}{\U{\mu m}}

\newcommand{\Cal}[1]{{\cal #1}}
\newcommand{\etal}{\textit{et al.}}
\newcommand{\U}[1]{\,{\rm{#1}}}

\newcommand{\X}[1]{_{\mathrm{#1}}}
\newcommand{\mat}[1]{\hbox{\boldmath{$#1$}\unboldmath}}

\newcommand{\euler}{\mathrm e}
\newcommand{\Sum}{\sum\limits}
\newcommand{\Prod}{\prod\limits}
\newcommand{\Int}{\int\limits}

\newcommand{\differential}{\>\mathrm d}

\newcommand{\E}[1]{\times 10^{#1}}

\newcommand{\XUV}{\textsc{xuv}}

\newcommand{\atopa}[2]{\genfrac{}{}{0pt}{}{#1}{#2}}

\newcommand{\nbh}{\hbox{-}}
\newcommand{\xray}{x\nbh{}ray}

\newcommand{\unitmatrix}{\mat{\mathbbm{1}}}

\newcommand{\tabFlourescenceNone}{I}

\newcommand{\tabFlourescenceNoneEnergy}{II}

\newcommand{\tabBindingNone}{III}

\newcommand{\tabAugerSingle}{IV}

\newcommand{\tabAugerSingleEnergy}{V}

\newcommand{\tabFlourescenceSingle}{VI}

\newcommand{\tabFlourescenceSingleEnergy}{VII}

\newcommand{\tabBindingSingle}{VIII}

\newcommand{\tabAugerDouble}{IX}

\newcommand{\tabAugerDoubleEnergy}{X}

\newcommand{\tabFlourescenceDouble}{XI}

\newcommand{\tabFlourescenceDoubleEnergy}{XII}

\newcommand{\tabBindingDouble}{XIII}

\documentclass[aip,jcp,10pt,twocolumn,showpacs,preprintnumbers,superscriptaddress,nolongbibliography]{revtex4-1}
\usepackage{bm,bbm,amsmath,amssymb,color,subeqnarray,CJK,graphicx}
\usepackage[nativepdf,bookmarks,bookmarksopen,bookmarksnumbered,raiselinks,breaklinks,%
pdftitle={Ultrafast absorption of intense x rays by nitrogen molecules},%
pdfauthor={Christian Buth, Ji-Cai Liu (刘纪彩), Mau Hsiung Chen (陳茂雄), James P. Cryan, Li Fang (方力), James M. Glownia, Matthias Hoener, Ryan N. Coffee, Nora Berrah},%
pdfsubject={Molecular physics},%
pdfkeywords={X rays, photoabsorption, Auger decay, light-matter interaction, nitrogen molecule, multiply-charged atoms, fragmentation, metastable dication, frustrated absorption}]{hyperref}

\begin{document}
\begin{CJK*}{UTF8}{}
\title{Ultrafast absorption of intense x~rays by nitrogen molecules}
\author{Christian Buth}
\thanks{Corresponding author}
\email{christian.buth@web.de}
\affiliation{Max-Planck-Institut f\"ur Kernphysik,
Saupfercheckweg~1, 69117~Heidelberg, Germany}
\affiliation{The PULSE Institute for Ultrafast Energy Science,
SLAC National Accelerator Laboratory, Menlo Park, California~94025, USA}
\affiliation{Department of Physics and Astronomy, Louisiana State
University, Baton Rouge, Louisiana~70803, USA}
\affiliation{Argonne National Laboratory, Argonne, Illinois~60439, USA}
\author{Ji-Cai Liu ({\CJKfamily{gbsn}刘纪彩})}
\affiliation{Max-Planck-Institut f\"ur Kernphysik,
Saupfercheckweg~1, 69117~Heidelberg, Germany}
\affiliation{Department of Mathematics and Physics, North China
Electric Power University, 102206~Beijing, China}
\author{Mau Hsiung Chen ({\CJKfamily{bsmi}陳茂雄})}
\affiliation{Physics Division, Lawrence Livermore National Laboratory,
Livermore, California~94550, USA}
\author{James P. Cryan}
\affiliation{The PULSE Institute for Ultrafast Energy Science,
SLAC National Accelerator Laboratory, Menlo Park, California~94025, USA}
\affiliation{Department of Physics, Stanford University, Stanford,
California 94305, USA}
\author{\hbox{Li Fang ({\CJKfamily{gbsn}方力})}}
\affiliation{Department of Physics, Western Michigan University, Kalamazoo,
Michigan~49008, USA}
\author{James M. Glownia}
\affiliation{The PULSE Institute for Ultrafast Energy Science,
SLAC National Accelerator Laboratory, Menlo Park, California~94025, USA}
\affiliation{Department of Applied Physics, Stanford University, Stanford,
California 94305, USA}
\author{Matthias Hoener}
\affiliation{Department of Physics, Western Michigan University, Kalamazoo,
Michigan~49008, USA}
\author{Ryan N. Coffee}
\affiliation{The PULSE Institute for Ultrafast Energy Science,
SLAC National Accelerator Laboratory, Menlo Park, California~94025, USA}
\affiliation{The Linac Coherent Light Source, SLAC National Accelerator
Laboratory, Menlo Park, California 94025, USA}
\author{Nora Berrah}
\affiliation{Department of Physics, Western Michigan University, Kalamazoo,
Michigan~49008, USA}
\date{31 May 2012}

\begin{abstract}
We devise a theoretical description for the response of nitrogen
molecules~(N$_2$) to ultrashort and intense x~rays from the free electron
laser~(FEL) Linac Coherent Light Source~(LCLS).
We set out from a rate-equation description for the \xray~absorption
by a nitrogen atom.
The equations are formulated using all one-\xray-photon
absorption cross sections and the Auger and
radiative decay widths of multiply-ionized nitrogen atoms.
Cross sections are obtained with a one-electron theory
and decay widths are determined from \emph{ab initio}
computations using the Dirac-Hartree-Slater~(DHS) method.
We also calculate all binding and transition energies of
nitrogen atoms in all charge states with the DHS~method as the
difference of two self-consistent field calculations~($\Delta$SCF~method).
To describe the interaction with~N$_2$,
a detailed investigation of intense \xray-induced ionization and
molecular fragmentation are carried out.
As a figure of merit, we calculate ion yields and the average charge state
measured in recent experiments at the LCLS.
We use a series of phenomenological models of increasing
sophistication to unravel the mechanisms of the interaction of
x~rays with~N$_2$: a single atom, a symmetric-sharing model,
and a fragmentation-matrix model are developed.
The role of the formation and decay of single and double core holes,
the metastable states of~N$_2^{2+}$, and molecular fragmentation are
explained.
\end{abstract}

%
%
%
%
%
%

\pacs{33.80.-b, 33.80.Eh, 32.80.Aa, 41.60.Cr}
\preprint{arXiv:1201.1896}
\maketitle
\end{CJK*}

\definecolor{mablack}{rgb}{0,0,0}
\definecolor{mared}{rgb}{1,0,0}
\definecolor{magreen}{rgb}{0,1,0}
\definecolor{mablue}{rgb}{0,0,1}
\definecolor{mamagenta}{rgb}{1,0,1}
\definecolor{macyan}{rgb}{0,1,1}
\definecolor{maorange}{rgb}{1,0.5,0}
\definecolor{mapurple}{rgb}{0.5,0,0.5}

\section{Introduction}

The theory of the interaction of ultrashort and intense x~rays with
atoms and molecules has recently come into focus by the newly-built
\xray~free electron laser~(FEL) Linac Coherent Light Source~(LCLS)
in Menlo Park, California, USA.~\cite{LCLS:CDR-02,Emma:FL-10}
The first experiments clearly revealed how important a detailed
understanding of the interaction of x~rays with single atoms and
small molecules is for their interpretation.~\cite{Young:FE-10,Hoener:FA-10,%
Cryan:AE-10,Fang:DC-10,Glownia:TR-10,Hau-Riege:NU-10,Doumy:NA-11,Cryan:AE-12}
To obtain a clear theoretical view on the physics of the
interaction with such comparatively simple systems represents
a crucial piece of basic research;
it is essential to form a foundation for studying the interaction
of more complex systems like condensed matter or biomolecules with
intense x~rays.
Due to the novelty of \xray~FELs, there is currently little information
available.
Naively, one may suspect that the impact of x~rays on matter depends only
on the absorbed dose, \ie, the energy deposited per unit mass of medium.
This would mean that the detailed pulse characteristics,
\ie, duration, spatial profile, and temporal shape, are irrelevant.
However, first theoretical and experimental studies of neon atoms
give a lucid explanation why this is too
simple a picture.~\cite{Rohringer:XR-07,Young:FE-10,Doumy:NA-11}
In this work, we report on the theoretical foundation of
the interaction of intense x~rays with small molecules and particularly
emphasize the additional challenges one faces over describing
interactions with single atoms due to molecular fragmentation.

\begin{figure}
  \begin{center}
    \includegraphics[clip,width=\hsize]{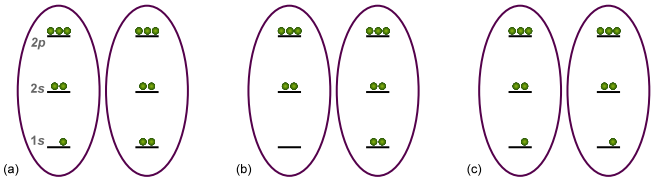}
    \caption{(Color online) Classification of key types of core
              holes in~N$_2$ in the localized core
              hole picture:~\cite{Cederbaum:DV-86,Tashiro:MD-10}
              (a)~a singly-occupied spatial core orbital on one of the atoms
              of~N$_2$: a single core hole~(SCH),
              (b)~a fully-depleted spatial core orbital on one N~atom: a
              double core hole~(DCH) on a single site~(ssDCH), and
              (c)~a singly-occupied spatial core orbital on both N~atoms: a
              DCH on two sites~(tsDCH).}
    \label{fig:SCH_DCH}
  \end{center}
\end{figure}

One of the simplest molecules, nitrogen~(N$_2$),
was studied extensively in the first experiments at
LCLS~\cite{Hoener:FA-10,Cryan:AE-10,Fang:DC-10,Glownia:TR-10,Hau-Riege:NU-10,%
Cryan:AE-12}
and the multiphoton nature of the interaction was characterized.
Specifically, Hoener~\etal~\cite{Hoener:FA-10} examined the interaction
of \xray~pulses of constant energy but with varying duration
and found in the ion yields the effect of frustrated \xray~absorption.
Next, Fang~\etal{}~\cite{Fang:DC-10} investigated the formation of
(multiple) core holes;
the most important core-hole types were found to be single core holes~(SCHs),
double core holes~(DCHs) on a single site~(ssDCH) and on two
sites~(tsDCH);~\cite{Cederbaum:DV-86,Tashiro:MD-10}
all of them are depicted in Fig.~\ref{fig:SCH_DCH}.
The tsDCH is a novel type of DCH which can be produced at third-generation
synchrotrons only with very low probability;~\cite{Lablanquie:ES-11}
its abundant production has only become feasible with the emerging
\xray~FELs.~\cite{Fang:DC-10,Cryan:AE-10,Cryan:AE-12,Berrah:DC-11}
Further, Cryan~\etal{}~\cite{Cryan:AE-10,Cryan:AE-12} investigated the
formation and decay of~DCHs in laser aligned~N$_2$ and the energy shift
of~DCHs with respect to the main SCH line where the synchronization between
the optical laser and LCLS was achieved with the method of
Glownia~\etal{};~\cite{Glownia:TR-10}
they found a clear sign of a nonspherical angular distribution of
the Auger electrons in the molecular reference frame.
Specifically, the experiments of Fang~\etal{} and Cryan~\etal{}
open up novel perspectives for chemical
analysis.~\cite{Cederbaum:DV-86,Santra:XT-09,Tashiro:MD-10,Berrah:DC-11}
Finally, Hau-Riege~\etal{}~\cite{Hau-Riege:NU-10} investigated
near-ultraviolet luminescence of~N$_2$.

The prototypical N$_2$~molecule has also been studied extensively
with other \xray/\XUV{}~sources.
Low-flux x~rays from third-generation synchrotrons have been
used to study the inner-shell properties
of~N$_2$~\cite{Stolte:DP-98,Ueda:HR-09} and
the dissociative one-\xray-photon absorption cross section was
measured.~\cite{Stolte:DP-98}
The production of SCHs by \xray~absorption results in
Auger electron spectra from SCH decay that were classified
in Refs.~\onlinecite{Siegbahn:FM-69,Stalherm:DI-69,Moddeman:DA-71,%
Agren:OI-81,Wetmore:TO-86}.
For photon energies below the SCH-formation threshold, we note
recent investigations at the Free Electron Laser
in Hamburg~(FLASH).~\cite{Jiang:FP-09,Jiang:TC-10}
The study of intense \XUV{}~absorption by~N$_2$ at FLASH has a
different character from \xray~absorption by~N$_2$ at LCLS as the former
radiation primarily targets valence electrons whereas
the latter almost exclusively interacts with core electrons.

In contrast to the one-photon physics at present synchrotrons,
the interaction of the x~rays from FELs with molecules is nonlinear.
Namely, for intense x~rays, several one-photon absorption events are
linked by the decay widths of the created holes.
So not only two or more photons are absorbed but the result of
the absorption also depends on the time-difference between the individual
one-photon absorption events.
If it did not depend on the time-difference, the outcome of
multiphoton absorption would only depend on the number of
absorbed photons and not on the timing of the processes, \ie,
the processes needed not be considered as an entity;
however, due to the finite lifetime of inner-shell holes, the two events
are linked and cannot be treated separately.

In this paper, we analyze theoretically the interaction
of intense x~rays with~N$_2$.
They represent the next step from studying the impact of x~rays
on single atoms like neon.~\cite{Rohringer:XR-07,Young:FE-10,Doumy:NA-11}
Molecules in intense \xray~FEL light of high enough
photon energy can be multiply core-ionized by sequentially absorbing
several \xray~photons.
Direct valence ionization occurs only rarely because the
associated cross sections are orders of magnitude smaller
than cross sections of core
electrons.~\cite{Als-Nielsen:EM-01,Thompson:XR-09,Buth:SU-up}
In very intense x~rays, this may even lead to a hollowing out of molecules,
\ie, core shells may be completely depleted due to~ssDCH formation.
For depleted shells, further ionization is drastically
reduced and can even lead to transparency for x~rays.~\cite{Nagler:TA-09}
Significant ionization of hollowed out molecules resumes only when Auger decay
refills the core shells.
We will theoretically analyze the copiousness and influence of core holes
for~N$_2$ in intense x~rays.
We set out with an investigation of isolated N~atoms and compute ion yields
and the average charge state.
Then, we consider N$_2$~molecules for which we have another
possibility that has significant impact on observables;
namely, as long as the molecular ion can be regarded as
an entity, \ie, dissociation has not yet separated the
molecule into fragments, the valence electrons are shared by both N~atoms.
This leads to a substantial redistribution of charge in decay processes
of core-ionized
molecules~\cite{Siegbahn:FM-69,Stalherm:DI-69,Moddeman:DA-71,Agren:OI-81,%
Wetmore:TO-86,Stolte:DP-98} which has a significant influence on the
observed ion yields from~N$_2$ (and many other) quantities.
Dicationic states~N$_2^{2+}$ are predominantly populated by Auger
decay of SCHs;
they are frequently metastable with respect to dissociation
with comparatively long lifetimes.~\cite{Wetmore:TO-86,Beylerian:CE-04}
In this case, dissociation on a rapid timescale only resumes
when more electrons are removed from the dication.
This interdependence of multiple \xray~absorption and nuclear
dynamics---which eventually leads to molecular fragmentation---poses
a challenge to theoretical molecular physics.
In the present study, we use a series of phenomenological models
to unravel the complicated electronic and nuclear processes;
it complements the phenomenological molecular rate-equation
model~\cite{footnote1} described in Ref.~\onlinecite{Hoener:FA-10}.

The paper is structured as follows: we present a theoretical study of
the absorption by~N$_2$ of ultrashort and intense \xray~pulses from
LCLS at a wavelength of~$1.1 \U{nm}$ ($1100 \eV$~photon energy)
with a nominal pulse energy of~$0.15 \U{mJ}$ for FWHM~pulse durations
of~$4 \U{fs}$ and a pulse energy of~$0.26 \U{mJ}$ for FWHM~pulse durations
of~$7$, $80$, and $280 \U{fs}$.
For multiply-ionized N~atoms, we use \emph{ab initio} computations to
determine Auger and fluorescence decay widths and the corresponding
transition energies, electron binding energies,
and one-\xray-photon absorption cross sections [Sec.~\ref{sec:elpropN}].
We formulate a rate-equation model for an N~atom
and study the sequential absorption of multiple \xray~photons in
Sec.~\ref{sec:yiatom}.
Ion yields of a single N~atom are predicted and show a strong deviation
from the experimentally measured ion yields of~N$_2$.
The atomic rate-equations are, therefore, extended phenomenologically
in Sec.~\ref{sec:yimolecule} to treat molecular fragmentation
which is the cause for the observed differences between the ion yields
of atoms and of molecules.
For this purpose, a series of models of increasing sophistication
is developed.
In Sec.~\ref{sec:micromodel}, we explain the phenomenology of~N$_2$
in intense x~rays in terms of elemental molecular processes.
Conclusions are drawn in Sec.~\ref{sec:conclusion}.

Our equations are formulated in atomic units.~\cite{Szabo:MQC-89}
For the conversion of a decay width~$\Gamma$ in electronvolts to a
lifetime~$\tau$ in femtoseconds, we use the
relation~$\tau = \frac{\hbar}{\Gamma} = \frac{0.658212 \U{eV \, fs}}{\Gamma}$.
All details of the models and calculations of this paper are
provided in the supplementary materials.~\cite{Buth:SU-up}

\section{Energies and decay widths of multiply-ionized nitrogen atoms}
\label{sec:elpropN}

In this section, we study the energy levels of
multiply-ionized N~atoms and their decay widths
and transition energies of Auger and radiative decay channels.
They will be used to describe the interaction of atoms
with x~rays in rate-equation approximation in Sec.~\ref{sec:yiatom}.
From the energy levels of multiply-charged N~atoms, we identify the
energetically accessible ionization channels for a specific \xray~photon
energy because only those electrons are available for direct
one-\xray-photon ionization that have a lower binding energy
than the photon energy.
The transition energies for Auger decay and \xray~fluorescence
allow an identification of the relevant processes in experimental electron
and photon spectra.
For elements from the second row of the periodic table, like nitrogen, only
Auger decay is important, if a competing Auger decay channel exists
in addition to \xray~fluorescence [Tables~\tabAugerSingle,
\tabFlourescenceSingle{} and \tabAugerDouble, \tabFlourescenceDouble{}
in Ref.~\onlinecite{Buth:SU-up}].

Energy levels, transition energies, and decay widths
for multiply-ionized N~atoms are calculated using the
Dirac-Hartree-Slater~(DHS) method.~\cite{Huang:NA-76,Chen:RL-79}
In the energy calculations, we use the DHS wave functions from the
appropriate defect configurations to compute the expectation values
of the total Hamiltonian including the Breit interaction,~\cite{Huang:NA-76}
\ie, a first-order correction to the local approximation is made.
We also include the effects of relaxation by performing separate
self-consistent field~(SCF) calculations for the initial and final states
[$\Delta$SCF~method] and take the difference of the total energies to obtain
binding energies~\cite{Huang:NA-76,Chen:RL-85} or
transition energies, respectively.
We calculate Auger and radiative decay widths with perturbation
theory~\cite{Grant:GI-74,Chen:RL-79} from frozen-orbital and
active-electron approximations as opposed to the relaxed-orbital
calculations used to determine energies.
The Auger matrix elements are calculated with DHS
wave functions for the initial state with a proper defect configuration.
Continuum wave functions are obtained by solving the Dirac-Slater
equations with the same atomic potential as for the initial state.
With this treatment, the orthogonality of the wave functions is assured.
The radiative decay widths are calculated
in multipole expansion using the relativistic length gauge for
the electric multipoles.~\cite{Grant:GI-74}
The radial matrix elements are also calculated using DHS wave functions
from the initial state.
The effects of relaxation on the total $K$-shell Auger and radiative
decay widths are less than~10~\% for atoms with a $K$-shell
vacancy.~\cite{Scofield:EC-74}
For a few times ionized N~atoms, the effects are even smaller of the
order of only a few percent.

The relativistic calculations of energies and decay widths for a N~atom yield
values in the $jj$~coupling scheme, \ie, they are specified with respect to
electronic configurations which are labeled by~$1s^\ell \, 2s^m \,
2p_{1/2}^{n_{1/2}} \, 2p_{3/2}^{n_{3/2}}$.
Here, $0 \leq \ell, m, n_{1/2} \leq 2$, and $0 \leq n_{3/2} \leq 3$ are
the occupation numbers of the respective orbitals where
the subscripts~$1/2$ and $3/2$ denote the value of the total angular momentum
quantum number~$j$.~\cite{Condon:TA-35}
By basing our method on DHS~wave functions, we develop a more general
approach than previous investigations, \eg, Ref.~\onlinecite{Bhalla:FY-75},
by treating the atomic electronic structure in one-electron approximation
including relativistic effects.
The DHS~approach yields good accuracy throughout the periodic table
and facilitates to assess, and if necessary incorporate, such
effects.
For light elements like nitrogen, relativistic effects are small and
have only minor importance for our modeling.
Specifically, spin-orbit coupling is small which allows us to
determine energies and decay widths in $LS$~coupling---where the
electronic configurations are denoted by~$1s^\ell \, 2s^m \, 2p^n$
with~$n = n_{1/2} + n_{3/2}$---from the corresponding quantities
in $jj$~coupling scheme.
Scalar relativistic effects are, nonetheless, accounted for in this
procedure.

In the supplementary materials,~\cite{Buth:SU-up} we discuss details
of the probabilistic averaging of energies and decay widths
in the $jj$~coupling scheme and we provide following data
for a multiply-ionized nitrogen atom:
the electron binding energies are gathered in Tables~\tabBindingNone,
\tabBindingSingle, and \tabBindingDouble;
we list the Auger decay widths in Tables~\tabAugerSingle{} and
\tabAugerDouble{} with the corresponding transition energies provided in
Tables~\tabAugerSingleEnergy{} and \tabAugerDoubleEnergy;
the \xray~fluorescence decay widths are presented
in Tables~\tabFlourescenceNone, \tabFlourescenceSingle,
and \tabFlourescenceDouble{} with the
respective transition energies collected in
Tables~\tabFlourescenceNoneEnergy, \tabFlourescenceSingleEnergy,
and \tabFlourescenceDoubleEnergy.
Our Auger decay widths for SCHs in Table~\tabAugerSingle{} are
in reasonable agreement with the early computations~\cite{Bhalla:FY-75} of
Bhalla.~\footnote{Only single core holes were considered by
Bhalla.~\cite{Bhalla:FY-75}
His results need to be averaged over the transitions
between multiplets in $LS$~coupling scheme to obtain values
for electronic configurations.}

\section{Ion yields of a nitrogen atom}
\label{sec:yiatom}

To describe the time-evolution of the absorption of x~rays and
the induced decay processes by a single N~atom, we use rate equations.
This approach has proven useful in many problems of light-matter
interaction and also for the interaction of x~rays from free electron lasers
with atoms.~\cite{Crance:DS-85,Rohringer:XR-07,Makris:MM-09,Milonni:LP-10}
The justifications for this approximation are,
first, that the envelope of the \xray~pulse varies slowly with respect
to the change of populations of states, \ie, ionization and decay processes
happen almost instantaneously on this time scale and,
second, coherence effects are small;
the x~rays induce essentially only one-photon absorption processes
which are, however, linked by inner-shell hole lifetimes.

The rate equations for a single nitrogen atom~\cite{Buth:SU-up} comprise
only electron removal terms and no electron addition terms
because only the former are relevant for our study and the
considered nitrogen charge states are always neutral or cationic.
We include all possible one-\xray-photon absorption processes
in the independent electron approximation;
the cross sections are obtained in one-electron approximation
with the \textit{Los Alamos National Laboratory Atomic Physics
Codes}.~\cite{Cowan:TA-81,LANL:AP-00}
All Auger decay widths of Sec.~\ref{sec:elpropN} are used,
however, \xray~fluorescence is only accounted for when there is no competing
Auger channel because Auger widths are typically orders of magnitude
larger than the corresponding fluorescence widths.~\cite{Buth:SU-up}
Other contributions are neglected, specifically, cross sections
for the simultaneous absorption of two x~rays are
tiny.~\cite{Doumy:NA-11}
Likewise more complicated and less important processes like
shake~off~\cite{Doumy:NA-11} and double Auger decay~\cite{Aberg:TP-75}
are not incorporated.

\begin{figure*}
  \begin{center}
    \hfill (a)~\includegraphics[clip,width=7cm]{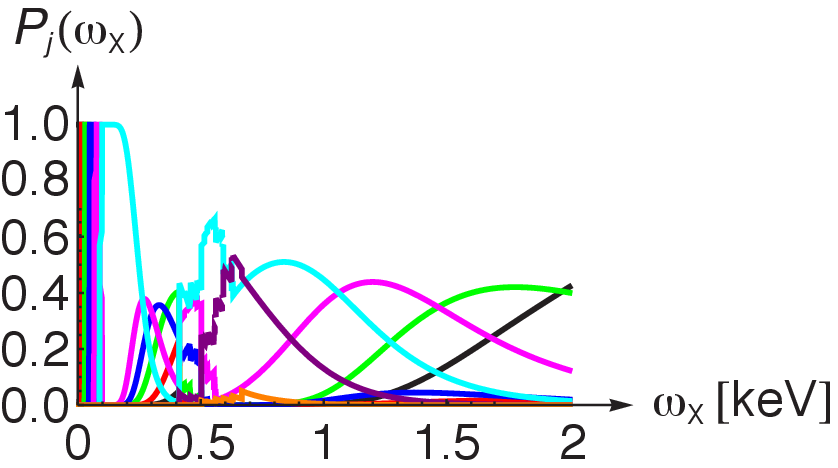} \hfill
    (b)~\includegraphics[clip,width=7cm]{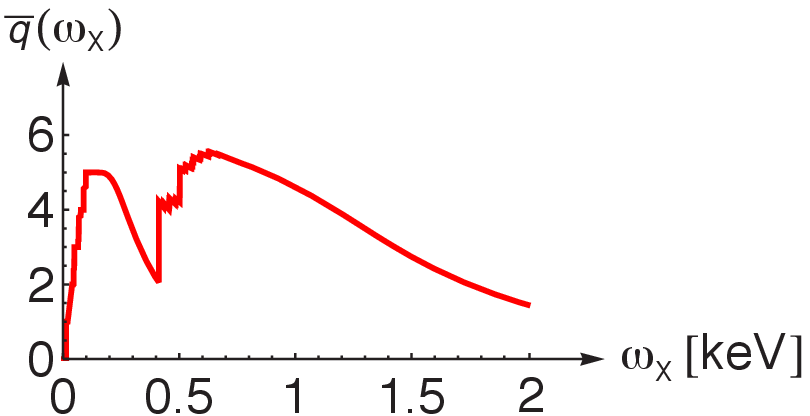} \hfill{}
    \caption{(Color) An N~atom after irradiation by x~rays of constant
             spatial beam profile with a photon fluence (photon flux
             integrated over all times) of~$8 \E{11} \, \U{\mu m^{-2}}$ and
             a Gaussian temporal shape~(\ref{eq:XrayGaussEnv}) with
             FWHM duration of~$2 \U{fs}$ at varying photon
             energy~$\omega\X{X}$.
             (a)~Probabilities~$P_j(\omega\X{X})$ from Eq.~(\ref{eq:chrgprob})
             to find an N~atom in specific charge states:
             neutral (\textbf{\textcolor{mablack}{black}}),
             singly (\textbf{\textcolor{mared}{red}}),
             doubly (\textbf{\textcolor{magreen}{green}}),
             triply (\textbf{\textcolor{mablue}{blue}}),
             quadruply (\textbf{\textcolor{mamagenta}{magenta}}),
             quintuply (\textbf{\textcolor{macyan}{cyan}}),
             sextuply (\textbf{\textcolor{mapurple}{purple}}), and
             septuply (\textbf{\textcolor{maorange}{orange}}).
             (b)~Average charge state~$\bar q(\omega\X{X})$ from
             Eq.~(\ref{eq:avcharge}).}
    \label{fig:yieldsphoton}
  \end{center}
\end{figure*}

The probability to find an N~atom at time~$t$ in an electronic
configuration~$1s^\ell \, 2s^m \, 2p^n$ is denoted by~$P_{\ell mn}(t)$
where~$\ell = m = 2$, $n = 3$ stands for the neutral atom.
The \xray~photon flux at time~$t$ is given by~$J\X{X}(t)$, \ie, the number of
\xray~photons that pass a unit area perpendicular to the beam in unit
time.~\cite{Als-Nielsen:EM-01}
The cross section at a chosen \xray~photon energy for the ionization
of an electron of an N~atom with electronic
configuration~$1s^\ell \, 2s^m \, 2p^n$ is denoted
as~$\sigma_{\ell mn}$ with~$0 \leq \ell, m \leq 2 \, \land \,
0 \leq n \leq 3$.
With these quantities, we formulate the rate equation to describe the
depletion of the ground state of nitrogen due to \xray~absorption:
\begin{equation}
  \label{eq:groundreq}
  \frac{\differential P_{223}(t)}{\differential t} =
    -\sigma_{223} \, P_{223}(t) \, J\X{X}(t) \; .
\end{equation}
The total (Auger plus \xray~fluorescence) decay width of a singly
core-ionized state~$1s^1 \, 2s^2 \, 2p^3$ is specified as~$\Gamma_{123}$.
The $1s$~ionization cross section of an N~atom
is denoted by~$\sigma_{123 \leftarrow 223}$.
Using these parameters, we find for the single core-hole rate
\begin{equation}
  \label{eq:singlereq}
  \begin{array}{rcl}
    \dfrac{\differential P_{123}(t)}{\differential t} &=& \bigl(
      \sigma_{123 \leftarrow 223} \, P_{223}(t) - \sigma_{123} \,
      P_{123}(t) \bigr) \, J\X{X}(t) \\
    &&{} - \Gamma_{123} \, P_{123}(t) \;  ,
  \end{array}
\end{equation}
and analogously for the double core-hole rate
\begin{equation}
  \label{eq:doublereq}
  \begin{array}{rcl}
    \dfrac{\differential P_{023}(t)}{\differential t} &=& \bigl(
      \sigma_{023 \leftarrow 123} \, P_{123}(t) - \sigma_{023} \,
      P_{023}(t) \bigr) \, J\X{X}(t) \\
    &&{} - \Gamma_{023} \, P_{023}(t) \; .
  \end{array}
\end{equation}
The three rate equations~(\ref{eq:groundreq}),
(\ref{eq:singlereq}), and (\ref{eq:doublereq}) already are sufficient
to describe the formation of~SCHs and DCHs in an N~atom.
They form a system of ordinary linear differential equations which is
solved numerically with \textit{Mathematica}~\cite{Mathematica:pgm-V8.0}
for the following initial conditions:
before the \xray~pulse, the N~atom is in its electronic
ground state, \ie, $P_{223}(-\infty) = 1$, and all other charge
states are unpopulated, \ie, $P_{\ell mn}(-\infty) = 0$
for~$\neg (\ell = m = 2 \, \land \, n = 3)$.
Many more rate equations---36~in total---are needed to describe
the ionization and various decay channels for all charge states from a
neutral to a septuply-ionized N~atom.

With the rate-equation description at hand, we can compute observables.
The probability to find an initially neutral N~atom in a cationic
state with charge~$0 \leq j \leq 7$ after the \xray~pulse is over
($t \to \infty$) is
\begin{equation}
  \label{eq:chrgprob}
  P_j = \sum_{\atopa{\scriptstyle \ell + m + n = 7 - j}{\scriptstyle
    0 \leq \ell, m \leq 2 \, \land \, 0 \leq n \leq 3}}
    P_{\ell mn}(\infty) \; .
\end{equation}
We define atomic ion yields to be normalized charge-state probabilities
with~$1 \leq j \leq 7$ by
\begin{equation}
  \label{eq:ionyield}
  Y_j =  \dfrac{P_j} {1 - P_0} \; ,
\end{equation}
where $1 - P_0 = \Sum_{k=1}^7 P_j$ is the probability to create a cation.
They are experimentally accessible and are a meeting point with
theory.~\cite{Young:FE-10,Hoener:FA-10}
The~$Y_j$ can be used to determine an average charge state~$\bar q$ via
\begin{equation}
  \label{eq:avcharge}
  \bar q = \Sum_{j = 1}^7 j \> Y_j \; .
\end{equation}
It is a useful gross quantity to characterize the \xray~induced ionization and
quantifies the average amount of radiation damage caused.~\cite{Hoener:FA-10}

Exemplary charge-state probabilities~(\ref{eq:chrgprob}) for an
isolated N~atom irradiated with intense x~rays are shown
in Fig.~\ref{fig:yieldsphoton}a
for a range of photon energies from~$0$ to~$2000 \eV$.~\cite{Buth:SU-up}
At low photon energies, certain ionization channels and, therefore,
higher charge states are not accessible because the photon energy is
not large enough to ionize inner electronic shells.
Specifically, energies of shells move to higher binding energies
with increasing charge state of the ions such that they move out of
reach for a one-photon ionization process with fixed \xray~photon
energy.~\cite{Rohringer:XR-07}
This is particularly clearly discernible in the range from~$400$
to~$700 \eV$ where steps indicate openings of ionization channels.
Above~$1200 \eV$, higher charge states decrease monotonically and the
ground-state remains more and more populated because the cross
sections have dropped substantially in magnitude such that the x~rays
do not ionize as effectively as before anymore.

The average charge state~$\bar q$ [Eq.~(\ref{eq:avcharge})] for a single
N~atom in intense x~rays is plotted in Fig.~\ref{fig:yieldsphoton}b
for the same range of photon energies as in Fig.~\ref{fig:yieldsphoton}a.
The above behavior of the charge-state probabilities~(\ref{eq:chrgprob})
is mirrored in the dependence of~$\bar q$ on the \xray~photon energy.
For low photon energies, $\bar q$~rises quickly due to large cross sections
but its maximum value is limited by the energetic accessibility of
electronic shells;
it peaks, first, around~$100 \eV$ when the valence shells of N~atoms
can be fully depleted (Table~\tabBindingNone{} in
Ref.~\onlinecite{Buth:SU-up}) producing quintuply
ionized final states and thus~$\bar q = 5$.
For higher photon energies $\bar q$~drops again due to decreasing
cross sections.
Starting at about~$400 \eV$ also core electrons of nitrogen can be
accessed leading to a second peak at about~$700 \eV$
when all electrons can be ionized.
This peak does not reach the maximum value of~$\bar q = 7$ because the
\xray~fluence is too low to fully strip an N~atom of its electrons.
This is in contrast to the first peak where the maximum value is actually
reached.
Above $700 \eV$~photon energy, $\bar q$~drops monotonically due
to decreasing cross sections.

To describe the interaction of finite pulses of limited transversal
extend---along the $x$~and $y$~coordinates with respect to the propagation
direction along the $z$~axis---with an ensemble of atoms,
we need to integrate over the temporal and spatial profiles
of the incident \xray~pulse in the focal region.
The LCLS is a FEL which operates by the self-amplification of
spontaneous emission~(SASE)
principle~\cite{Kondratenko:GC-79,Bonifacio:CI-84,Saldin:PF-00}
which produces chaotic radiation pulses with random spikes.
In principle, the pulse structure of SASE pulses needs to be taken into
account in nonlinear optical processes.
However, as exemplified for atoms in Ref.~\onlinecite{Rohringer:XR-07},
the impact of chaotic light on linked one-photon
processes---which are the predominant mode of interaction of
intense x~rays with nitrogen---has little impact.
Therefore, there is no need to treat the explicit form of SASE~pulses in this
problem [see also Table~\ref{tab:chprobs} for a comparison of results
for a Gaussian pulse with results from SASE pulses].
Instead, it is sufficient to use a Gaussian temporal profile for
the rate of photons
\begin{equation}
  \label{eq:XrayGaussEnv}
  \Gamma\X{X}(t) = \Gamma\X{X,0} \> \euler^{-4 \ln 2 \, \textstyle{(}
    \frac{t}{\tau\X{X}} \textstyle{)}^2} \; ,
\end{equation}
with a peak rate of~$\Gamma\X{X,0} = 2 \, \sqrt{\frac{\ln 2}{\pi}} \>
\frac{n\X{ph}}{\tau\X{X}}$ for~$n\X{ph}$~photons in the pulse and
a full width at half maximum~(FWHM) duration of~$\tau\X{X}$.
The number of photons~$n\X{ph}$ is linked to the total energy in the
pulse~$E\X{P}$ via~$n\X{ph} = \frac{E\X{P}}{\omega\X{X}}$ for
approximately monochromatic x~rays with photon energy~$\omega\X{X}$.
The LCLS beam is modeled as a Gaussian beam~\cite{Milonni:LP-10}
with a long Rayleigh range which implies that the spatial profile of the
beam along its axis, the $z$~axis, does not vary appreciably over
the acceptance volume of the ion spectrometers.
Therefore, the \xray~flux is approximately constant along the $z$~axis and
thus we may disregard its $z$~dependence in what follows.
There is, however, a substantial variation of the \xray~flux transversally
to the beam axis.
The transversal spatial profile of LCLS pulses was
derived from ablation of solids in the FEL
beam~\cite{Chalupsky:CF-07,Barty:PC-09,Hoener:FA-10,Chalupsky:CD-11,%
Moeller:PB-11} that yielded an elliptic profile with a FWHM length of
the major and minor axes of~$\varrho\X{x}
= 2.2 \mum$ by~$\varrho\X{y} = 1.2 \mum$, respectively.
We have the following analytic expression for the transverse beam profile
\begin{equation}
  \label{eq:beamprofile}
  \varrho(x,y) = \frac{4 \ln 2}{\pi \, \varrho\X{x} \,
    \varrho\X{y}} \,
    \euler^{-4 \ln 2 \, [(x / \varrho\X{x})^2 +
                         (y / \varrho\X{y})^2]} \; ;
\end{equation}
it is normalized to unity with respect to an integration over
the entire $x$--$y$~plane, \ie, $\Int_{-\infty}^{\infty}
\Int_{-\infty}^{\infty} \varrho(x,y) \, \differential x
\differential y = 1$.
In our computations we find that an integration in the $x$--$y$~plane
over an area of~$10^2 \U{\mu m}^2$ centered on the beam axis
yields converged ratios of DCHs versus SCHs.
Finally, the position-dependent \xray~flux follows from
Eqs.~(\ref{eq:XrayGaussEnv}) and (\ref{eq:beamprofile}) to
\begin{equation}
  \label{eq:XrayFlux}
  J'\X{X}(x,y,t) = \varrho(x,y) \> \Gamma\X{X}(t) \; .
\end{equation}
We use only a fraction of the nominal pulse
energy~\cite{Hoener:FA-10,Young:FE-10}
to determine~$n\X{ph}$ to account for losses in the experimental setup.
This percentage is determined in Sec.~\ref{sec:fragmat}
by a fit of experimental data with our theoretical predictions.

\section{Ion yields of a nitrogen molecule}
\label{sec:yimolecule}

\begin{figure*}
  \begin{center}
    (a) \includegraphics[clip,width=6cm]{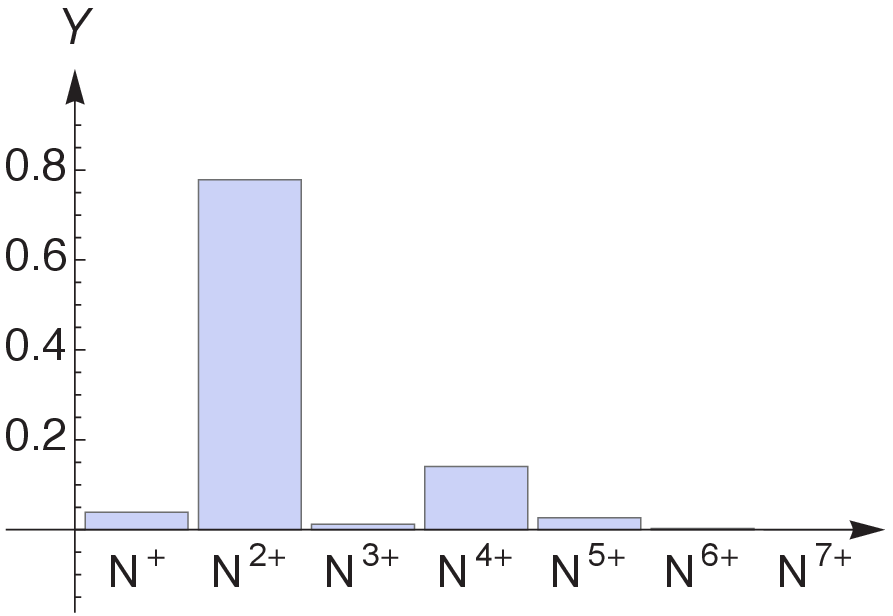}
    \quad
    (b) \includegraphics[clip,width=6cm]{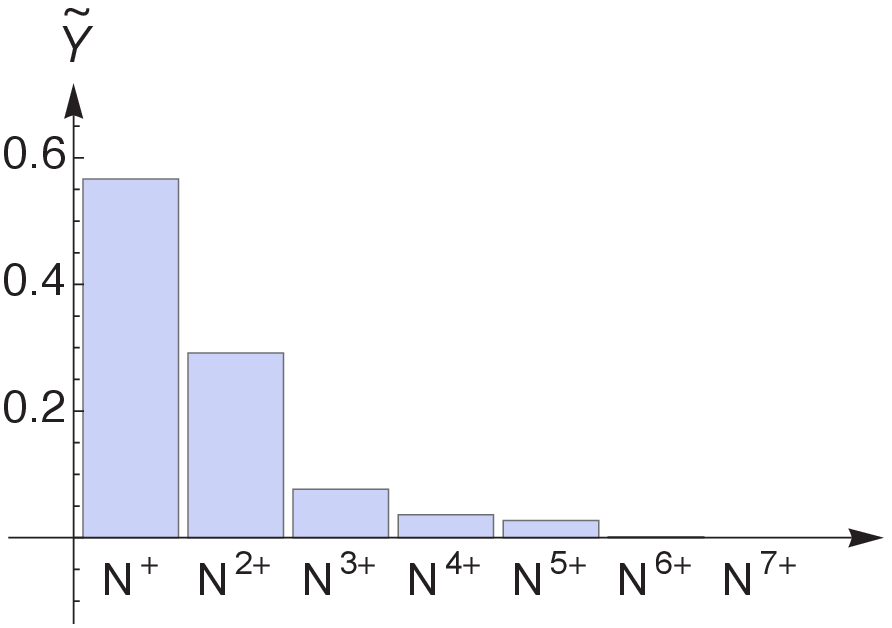}
    \caption{(Color online) Ion yields of (a)~an N~atom and (b)~an
             N$_2$~molecule irradiated by x~rays of a FWHM duration
             of~$280 \U{fs}$ and a photon energy of~$1100 \eV$.
             In (a)~we show theoretical results from the single atom model of
             Sec.~\ref{sec:yiatom} for the \xray~flux~(\ref{eq:XrayFlux}).
             We assume that
             only~$31 \%$ of the nominal pulse energy
             of~$0.26 \U{mJ}$ are actually available in the LCLS
             experiment [Table~\ref{tab:fragconsts}].
             In (b)~we show experimental data for an N$_2$~molecule.
             The comparison of atomic~$Y_j$ with molecular~$\tilde Y_j$
             ion yields for nitrogen clearly demonstrates the impact
             of molecular processes during and after the interaction
             with the x~rays.}
    \label{fig:yiatmmol}
  \end{center}
\end{figure*}

In the previous Sec.~\ref{sec:yiatom}, we discussed the ion yields of an
isolated N~atom which we denoted by~$Y_j$ for the positive
charge~$j \in \{1, \ldots, 7\}$.
In this section ion yields from an N$_2$~molecule---after fragmentation into
atomic ions---are calculated to which we reference by~$\tilde Y_j$.
Figure~\ref{fig:yiatmmol} shows the theoretical ion yields for an
N~atom alongside the experimental data for an N$_2$~molecule;
we find a stark difference between both bar charts.
The trend found in the ion yields for a specific set of parameters
solidifies, if we vary the pulse duration [Fig.~\ref{fig:avcharge}] where
we use~$\bar q$ [Eq.~(\ref{eq:avcharge})] as a figure of merit.

At first sight, one may assume that the absorbed dose~\cite{Young:FE-10}
and thus the ionization of a sample of~N$_2$ is very similar when the
\xray~pulses have the same total energy and central wavelength and only
one-photon absorption processes are included in the
rate equations [Sec.~\ref{sec:yimolecule}].
If there was only a single one-photon process with cross section~$\sigma_1$,
then the probability of ionization after the \xray~pulse would
only depend on the \xray~fluence, \ie, $\sigma_1 \Int_{-\infty}^{\infty}
J\X{X}(t) \differential t$, and thus be obviously independent of the
time evolution of the \xray~flux~$J\X{X}(t)$.
In Fig.~\ref{fig:avcharge}, we display~$\bar q$ with respect to the
\xray~pulse duration determined from the experimental data and for an N~atom.
One may expect naively that the absorbed dose is similar in all cases
and $\bar q$~is the same for all pulse durations, \ie, a horizontal
line in Fig.~\ref{fig:avcharge}.
However, the contrary is true;
we find a substantial dependence of~$\bar q$ on the pulse duration.
This finding is indicative of (nonlinear) multi-\xray-photon physics
in the sample, \ie, the absorption of several x~rays.

As expected from Fig.~\ref{fig:yiatmmol}, there is a large discrepancy
between~$\bar q$ for an N~atom and the experimentally observed
$\bar q$~for~N$_2$.
These differences \emph{cannot} be ascribed to molecular effects
on photoionization cross sections~\cite{Lucchese:MP-05}
because they are small for the chosen photon energies which are high above
the ionization thresholds of
multiply-charged~N$_2$.~\cite{Cohen:IP-66,Lucchese:MP-05,%
Gokhberg:PC-09,Ueda:HR-09}
Likewise fluorescence and Auger decay widths in molecules are usually
well-represented by the respective values of the isolated atoms.

The differences between the molecular ion
yields and the atomic ion yields in Fig.~\ref{fig:yiatmmol}
are caused by the interplay of core ionization of~N$_2$ with Auger decay,
and nuclear dynamics.
Namely, \xray~absorption by~N$_2$ predominantly leads to SCH
formation and N$_2^+$~cations;
unless more x~rays are absorbed shortly after, the SCH Auger decays producing
the molecular dication~N$_2^{2+}$.
The N$_2^{2+}$ fragments as follows:
\begin{subeqnarray}
  \label{eq:fragN2pp}
  \slabel{eq:fragNppN}
  \mathrm N_2^{2+} && \longrightarrow \mathrm N{\phantom{^+}} + \mathrm N^{2+} \\
  \slabel{eq:fragNpNp}
  \mathrm N_2^{2+} && \longrightarrow \mathrm N^+ + \mathrm N^+ \; ,
\end{subeqnarray}
where the probabilities for the channels~(\ref{eq:fragNppN}) and
(\ref{eq:fragNpNp})
%
%
are~$0.26$
%
%
and~$0.74$, respectively.%
~\footnote{%
The fragmentation probabilities for the two channels in Eq.~(\ref{eq:fragN2pp})
are deduced from the decay products and probabilities of
a SCH in~N$_2$ measured in experiments at third-generation synchrotrons.
Namely, the dissociative photoionization cross sections of N$_2$
for the production of~N$^+$, N$^{2+}$, N$^{3+}$, and N$_2^+$
by one-x-ray-photon absorption have been measured
%
%
in the photon-energy range from~100 to $800 \eV$.~\cite{Stolte:DP-98}
We use the functional relationship for photon energies above the
nitrogen $K$-edge in Table~A of Ref.~\onlinecite{Stolte:DP-98}
to extrapolate to a photon energy of~$1100 \eV$
%
%
yielding~$\sigma_{\mathrm N_2^+} = 1.83 \U{kbarn}$,
%
%
$\sigma_{\mathrm N^+} = 85.0 \U{kbarn}$, and
%
%
$\sigma_{\mathrm N^{2+}} = 30.4 \U{kbarn}$ as was used in
Refs.~\onlinecite{Hoener:FA-10,footnote1} to parametrize the molecular rate
equations.}
The N$_2^{2+}$ are an extreme case in the sense that
the impact of molecular fragmentation is very pronounced
on the ratio of N$^+$~yield versus N$^{2+}$~yield.
The existence of the two fragmentation channels~(\ref{eq:fragN2pp}) offers a
lucid explanation for the observed differences in Fig.~\ref{fig:yiatmmol}
between atomic and molecular ion yields.
For low \xray~flux, essentially no further x~rays are absorbed prior to
the Auger decay of a SCH in an N~atom, \ie, the initially neutral
atom ends up as~N$^{2+}$.
However, for Auger decay of a SCH in~N$_2$, the two charges,
may be shared between both atoms~(\ref{eq:fragNpNp}).
This fact explains the large amount of N$^{2+}$~ion yield for the atom
versus the large N$^+$~ion yield for the molecule in Fig.~\ref{fig:yiatmmol}.
A similar argument explains also the different ratio of the ion
yields of~N$^{3+}$ versus~N$^{4+}$;
namely, in the case that \xray~absorption is fairly slow---such that Auger
decay and fragmentation mostly occur before another x~ray is
absorbed---a second x~ray is more likely absorbed by
a~N$^+$ fragment than a N$^{2+}$~fragment as the former are more
abundant than the latter.
A subsequent Auger decay produces then preferably~N$^{3+}$
instead of~N$^{4+}$.

Despite the seemingly clear view on the physical reasons behind the
deviations between~$Y_j$ and~$\tilde Y_j$ in Fig.~\ref{fig:yiatmmol},
one should bear in mind that N$_2^{2+}$~posses metastable dicationic
states with respect to dissociation.
The lifetime of~N$_2^{2+}$ can extend even to a few
microseconds~\cite{Wetmore:TO-86} for the dissociation
through tunneling into the $\mathrm N^+ + \mathrm N^+$~continuum.
In a laser-dissociation experiment,~\cite{Beylerian:CE-04}
$80 \U{fs}$ and $120 \U{fs}$~for the~$\mathrm N^+ + \mathrm N^+$ and
$\mathrm N + \mathrm N^{2+}$~fragmentation channels, respectively,
were measured which are on a comparable time scale with
the LCLS pulse durations discussed here.
Consequently, the impact of the metastable dicationic states is to slow down
fragmentation with a time constant of~${\sim}100 \U{fs}$.
This assessment is also supported by the fact that less
than~$2 \%$ of the N$^+$~peak in the time-of-flight spectra
were ascribed to undissociated~N$_2^{2+}$ in
Ref.~\onlinecite{Hoener:FA-10}.
For metastable states, dissociation on a rapid timescale may only resume
when more electrons are removed from the dication.
In the following Secs.~\ref{sec:lowerlim} and \ref{sec:fragmat},
we present several phenomenological models to account
for molecular fragmentation.

\subsection{Symmetric-sharing model}
\label{sec:lowerlim}

Here we develop the simplest model of breakup
of~N$_2$ after irradiation by x~rays in which charges produced by
\xray~absorption on both N~atoms are shared equally upon molecular
fragmentation for an even number of charges;
for an odd number of charges, one of the fragments carries one
excess charge with respect to the other fragment.
This model aims to contrast the extreme view of
the single-atom model for ion yields of~N$_2$ from Sec.~\ref{sec:yiatom}
in which \emph{no} redistribution of charge over the atoms
in a molecule is accounted for in the course of the fragmentation process.
Symmetric sharing is suggested in our situation by the dissociation
of~N$_2^{2+}$ [Eq.~(\ref{eq:fragN2pp})], the analysis
of the potential energy curves of~N$_2^{3+}$ which indicate a
most probable dissociation into~$\mathrm N^+ +
\mathrm N^{2+}$,~\cite{Bandrauk:ES-99} and laser-dissociation
experiments of~N$_2^{4+}$ for which symmetric sharing
was found to be the most prominent fragmentation
channel.~\cite{Codling:CS-91,Baldit:CE-05}
As a redistribution of charge---that is produced treating the atoms
independently---never increases the maximum charge found on the
involved atoms, the single-atom ion yields of Sec.~\ref{sec:yiatom} lead to an
upper limit of the average charge state~$\tilde{\bar q}$
from~N$_2$.
Here we investigate the other extreme position which is to assume
that the electrons of a charged N$_2$~molecule are always shared
equally between both N~atoms upon fragmentation.
In other words, the charge on the two atoms of~N$_2$ is determined
independently and then redistributed between the two
N~atoms.~\cite{Buth:SU-up}
Such an equal sharing of charges leads to the smallest amount of
positive charge per atom and thus to the lowest~$\tilde{\bar q}$.
Hence it represents a lower limit to the observed
charge-state distribution.

The~$\tilde{\bar q}$ from the symmetric-sharing model is displayed in
Fig.~\ref{fig:avcharge}.
Interestingly, we find that the lower limit is approached quite closely
by the experimental data for short x~ray pulses.
However, the slow progression of the lower limit of~$\tilde{\bar q}$
with the pulse duration leads to a significant deviation for longer pulses.
This indicates that, for short pulses, there is mostly equal sharing of the
charges between the atoms in~N$_2$ while there is
a different mechanism at work for longer pulses.
Of course, this is related to the fact that the importance of the various
processes entering the atomic rate equations of Sec.~\ref{sec:yiatom}
are weighted differently depending on the actual
LCLS pulse parameters.
As will be discussed in detail in Sec.~\ref{sec:micromodel}, the principal
reasons for the observed deviations is that
during the long \xray~pulses, Auger decay processes mostly occur prior
absorbing more x~rays and fragmentation may happen in between.
However, for short pulses, the molecule charges up quickly
and nuclear expansion and fragmentation are of little importance
on the time scale of the shortest ($4 \U{fs}$) pulse.
In this case, only after the pulse is over, there is sufficient time
for the induced nuclear wavepacket to evolve leading to
molecular breakup.

\subsection{Fragmentation-matrix model}
\label{sec:fragmat}

To account for fragmentation of~N$_2$ in the course of
the interaction with x~rays in a more realistic way,
we make the following heuristic ansatz.
We use the probabilities from the atomic rate equations after the
\xray~pulse is over to form probabilities to find N$_2$~in various
charge states treating the two N~atoms as independent.
The fragmentation of the molecular ion is then assumed to be determined
only by the charges on the two atoms in terms of fixed ratios
independent of the timing of the interaction with the x~rays.
This approximation is motivated by realizing that the fragmentation
of~N$_2^{2+}$---for which we have the two channels of
Eq.~(\ref{eq:fragN2pp})---depends only on the amount of~N$_2^{2+}$
produced in the course of the interaction with the \xray~pulse.
A fixed fragmentation constant then accounts for the redistribution
of charges.
Introducing constants for all relevant fragmentation patterns
of~N$_2$ in all possible charge states leads to a fragmentation matrix
which transforms the molecular probabilities from the independent atom
treatment into molecular probabilities that account for charge
redistribution from which ion yields are determined that
can be compared with experimental data.
Using a fixed fragmentation pattern clearly implies that we assume
that fragmentation always occurs in the same way for a given
independent-atom molecular charge state after the \xray~pulse is over, \ie,
we neglect all dependencies on the actually involved intermediate
states which change for different \xray~pulse parameters and dissociation
during the \xray~pulse and focus exclusively on the achieved
final independent-atom molecular charge state.

For a mathematical formulation of the fragmentation-matrix model, we
set out from the atomic probabilities to find an N~atom in
various charge states after the \xray~pulse is
over~$P_j$ with~$0 \leq j \leq 7$.
They are readily obtained from the solution of the atomic rate
equations of Sec.~\ref{sec:yiatom} via Eq.~(\ref{eq:chrgprob}).
Independent-atom molecular probabilities~$P'_{ij}$ of~N$_2$ for
finding charge state~$i$ on the left N~atom and charge state~$j$
on the right N~atom follow then immediately from~$P'_{ij}
= P_i \, P_j = P'_{ji}$ for~$0 \leq i, j \leq 7$;
they are aggregated in the vector~$\vec P'$.
The fragmentation matrix is denoted by~$\mat F$;
it transforms the~$\vec P'$ into molecular
probabilities that account for charge redistribution~$\vec {\tilde P}$ via
\begin{equation}
  \label{eq:fragprob}
  \vec {\tilde P} = \mat F \, \vec P' \; .
\end{equation}
The vector~$\vec {\tilde P}$ is formed by the molecular
probabilities~$\tilde P_{ij} = \tilde P_{ji}$ to find~N$_2^{(i+j)+}$.
\emph{Nota bene}, Eq.~(\ref{eq:fragprob}) needs to be applied for a single
\xray~intensity;
volume averaging~(\ref{eq:beamprofile}) is performed for the transformed result.
The atomic probabilities~$P_j$ [Eq.~(\ref{eq:chrgprob})] are recovered
from Eq.~(\ref{eq:fragprob}) by choosing~$\mat F = \unitmatrix$.
Summing over one index of~$\tilde P_{ij}$ then yields
\begin{equation}
  \label{eq:reduceprob}
  P_j = \Sum_{i=0}^7 {\tilde P_{ij}}
\end{equation}
because~$\Sum_{i=0}^7 P_i = 1$ holds.
Relation~(\ref{eq:reduceprob}) is a general property
of any~$\vec {\tilde P}$ from a meaningful~$\mat F$ to find the
probabilities~$P_j$ of atomic fragments after molecular breakup.
For an arbitrary~$\mat F$, we find by setting~$\vec P'$
in Eq.~(\ref{eq:fragprob}) to Cartesian unit vectors
that the sum of the matrix elements in each column of~$\mat F$ is unity, \ie,
the sum of the elements of a vector remains the same
(unity for probabilities) upon transformation with~$\mat F$.

To find a form of~$\mat F$ that introduces a redistribution
of charge, we make the additional assumption that a
molecular ion will dissociate into fragments
with similar charge states.
Specifically, we assume that a molecular ion with an even number of
electrons predominantly fragments symmetrically and, for an odd number
of electrons, it fragments such that only one excess charge is found
on one of the two fragments.
As discussed previously, the independent-atom view [Sec.~\ref{sec:yiatom}]
and the symmetric-sharing view [Sec.~\ref{sec:lowerlim}] are
two extreme positions with the former representing
the upper limit of the average charge state [Eq.~(\ref{eq:avcharge})]
and the latter being its lower limit.
Hence a redistribution of charge between the independent-atom
channel and the symmetric-sharing channel, with a certain amount
of probability shifted from the former to the latter,
is able to represent the impact on~$\tilde{\bar q}$
even if more fragmentation channels are present for a specific molecular
charge state.
In other words, we realize that only the coupling of~$P'_{ij}$
for~$|i-j| \geq 2$ to more symmetric charges states needs
to be considered for our purposes and a fraction~$0 \leq f_{\{i,j\}}
\leq 1$ of the probability of the
charge state~$\{i,j\}$ goes over into~$\{k,l\} = \{(i+j)/2,(i+j)/2\}$
for even~$i+j$
and into~$\{k,l\} = \{(i+j-1)/2,(i+j+1)/2\}$ for odd~$i+j$.%
\footnote{%
The notation in terms of sets indicates that the ordering of~$i$, $j$ and
$k$, $l$ is not relevant as~$\tilde P_{ij} = \tilde P_{ji}$.}
This kind of redistribution of probability can be introduced easily
using a transformation matrix~$\mat{\Cal F}_{\{i,j\}}$
which differs from the identity matrix
by setting~$(\mat{\Cal F}_{\{i,j\}})_{(i,j),(i,j)} =
(\mat{\Cal F}_{\{i,j\}})_{(j,i),(j,i)}
= 1-f_{\{i,j\}}$ and $(\mat{\Cal F}_{\{i,j\}})_{(k,l),(i,j)} =
(\mat{\Cal F}_{\{i,j\}})_{(l,k),(j,i)} = f_{\{i,j\}}$.
The set of 21~molecular charge states~$\{i,j\}$ for which~$|i-j| \geq 2$
holds is
\begin{eqnarray}
  \label{eq:fragcons}
  K &=& \{\{0,2\}, \{0,3\}, \{0,4\}, \{0,5\}, \{0,6\}, \{0,7\}, \nonumber \\
  && \hphantom{\{} \{1,3\}, \{1,4\}, \{1,5\}, \{1,6\}, \{1,7\}, \{2,4\}, \\
  && \hphantom{\{} \{2,5\}, \{2,6\}, \{2,7\}, \{3,5\}, \{3,6\}, \{3,7\},
    \nonumber \\
  && \hphantom{\{} \{4,6\}, \{4,7\}, \{5,7\}\} \; . \nonumber
\end{eqnarray}
For all~$k \in K$, we find a~$\mat{\Cal F}_k$ from which we construct
the fragmentation matrix via~$\mat F = \Prod_{k \in K} \mat{\Cal F}_k$.
As~$\mat{\Cal F}_k$ and $\mat{\Cal F}_l$ commute for all~$k, l \in K$ and
matrix multiplication is associative,
$\mat F$~does not depend on the sequence of multiplication of
the~$\mat{\Cal F}_k$.
The set~$K$ [Eq.~(\ref{eq:fragcons})] contains all necessary
molecular charge states to redistribute charges as is done
in the symmetric-sharing model of Sec.~\ref{sec:lowerlim}.
Consequently, the probabilities of Sec.~\ref{sec:lowerlim} are
recovered by choosing~$f_k = 1$ for all~$k \in K$
and using Eq.~(\ref{eq:reduceprob}).

\begin{table}
  \centering
  \begin{ruledtabular}
    \mbox{\begin{tabular}{cc|cccccc}
       $\tau\X{X}$ [$\mathrm{fs}$] & $E\X{P}$ [$\mathrm{mJ}$] &
       $f_{\{0,3\}}$ & $f_{\{0,4\}}$ & $f_{\{2,4\}}$
       & $f_{\{0,5\}}$ & $f_{\{2,5\}}$ & $f_{\{0,6\}}$ \\
    \hline
                 280 & $0.31 \times 0.26$ & 0 & 0.54 & 1 & 0 & 0 & 1  \\
       \phantom{0}80 & $0.25 \times 0.26$ & 0 & 0.61 & 1 & 0.26 & 0.17 & 1 \\
       \phantom{00}7 & $0.16 \times 0.26$ & 0 & 1 & 1 & 1 & 1 & 1 \\
       \phantom{00}4 & $0.26 \times 0.15$ & 1 & 1 & 0.48 & 1 & 0 & 0.65
    \end{tabular}}
  \end{ruledtabular}
  \caption{Fragmentation constants for the most probable channels
           for the four pulse durations.
           Here, $f_{\{0,2\}} = 0.74$~in all cases [Eq.~(\ref{eq:fragN2pp})].
           The LCLS FWHM pulse duration is~$\tau\X{X}$
           [Eq.~(\ref{eq:XrayGaussEnv})], the LCLS photon energy
           is~$1100 \eV$, and the actually available pulse energy in the LCLS
           experiment is~$E\X{P}$.
           A nominal pulse energy of~$0.15 \U{mJ}$ is specified
           for $4 \U{fs}$~pulses and $0.26 \U{mJ}$ for the remaining
           three pulse durations.}
  \label{tab:fragconsts}
\end{table}

Having asserted us that the fragmentation-matrix model correctly
integrates both the single-atom model and the symmetric-sharing model,
we need to find a good approximation for the 21~parameters
in~$\mat F$.
As insufficient theoretical information has been published on the
fragmentation of highly-charged~N$_2$, we need to find
the parameters with the help of experimental data.
The quest for the right parameters is further hindered by
the fact that the nominal pulse energies specified differ from the actually
available energies.~\cite{Young:FE-10,Hoener:FA-10}
The discrepancy between the nominal and the actual value of the pulse
energy at the sample in the experiment is
caused by losses in the beam line transmission which are not
particularly well determined.
In Ref.~\onlinecite{Hoener:FA-10}, it is specified that only
$15 \, \%$--$35 \, \%$~of the nominal pulse energy arrive at the sample.
In order to determine the fragmentation constants, we
fit the theoretical ion yields from the fragmentation-matrix
model [Eq.~(\ref{eq:ionyield})] to the experimental ion yields of~N$_2$
by minimizing the modulus of the difference between the
average charge states~(\ref{eq:avcharge}) with respect to
the fragmentation constants at the four measured \xray~pulse
durations, \ie, for~$4$, $7$, $80$, and~$280 \U{fs}$.~\cite{Buth:SU-up}
Thereby, we adjust the pulse energies such that the constant for the
fragmentation of~N$_2^{2+}$ [Eq.~(\ref{eq:fragN2pp})],
$f_{\{0,2\}} = 0.74$, \ie, the experimental value for
fragmentation after Auger decay of a SCH is recovered.
This procedure yields the fragmentation constants
and the fractions of the nominal pulse energy that
are listed in Table~\ref{tab:fragconsts}.
Note that this approach assumes that the molecular charge state~${\{0,2\}}$
is predominantly produced by SCH~decay, \ie, the competing
processes of double valence ionization by x~rays is
neglected because of tiny cross sections.

A number of insights about the experiment can be gained by
inspecting Table~\ref{tab:fragconsts}.
Examining~$E\X{P}$, we find that the actually
available pulse energy in the experiment is with one exception
in the range~$25$--$31 \%$ of the nominal pulse energy.
Only the pulse energy for~$7 \U{fs}$~pulses is around~$16 \%$
of the nominal pulse energy of~$0.26 \U{mJ}$.
This drop can be rationalized by taking~$28 \%$ of~$0.15 \U{mJ}$---the
nominal pulse energy specified for the case of~$4 \U{fs}$~pulses---which
leads to the same actual pulse energy as~$16 \%$ of~$0.26 \U{mJ}$.
The value~$28 \%$ is consistent with the percentages of
the other pulse durations and indicates that potentially there
was a drop in the actual output of LCLS when going to so short pulses.
The range~$25$--$31 \%$ is higher than the value of~${\sim}23 \%$
specified in Ref.~\onlinecite{Young:FE-10} for~$1050 \eV$~photon energy.
The peak \xray~intensities follow
%
%
to~$3.1 \E{17} \U{W / cm^2}$, $1.7\E{17} \U{W / cm^2}$,
$2.6 \E{16} \U{W / cm^2}$, and $9.0 \E{15} \U{W / cm^2}$
for the nominal pulse durations~$4$, $7$,
$80$, and $280 \U{fs}$, respectively.

The change of molecular fragmentation with pulse parameters can
be read off Table~\ref{tab:fragconsts}.
Predominantly, $f_{\{0,2\}}$~is determined by the absorption of
one x~ray, $f_{\{0,3\}}$ and $f_{\{0,4\}}$~are from two-\xray-absorption,
$f_{\{2,4\}}$, $f_{\{0,5\}}$, and $f_{\{0,6\}}$~are from three-\xray-absorption,
and $f_{\{2,5\}}$~is from four-\xray-absorption.
Apart from~$f_{\{2,4\}}$ and $f_{\{0,6\}}$ the fragmentation constants overall
increase with decreasing pulse duration.
This indicates that symmetric sharing becomes very important
for the shortest pulses while molecular fragmentation may occur during
the longer pulse prior absorption of further x~rays.
Due to very low probabilities, the values for~$f_{\{2,4\}}$,
$f_{\{2,5\}}$, and $f_{\{0,6\}}$ for $4 \U{fs}$~pulses are not important.
As~$f_{\{2,4\}}$ and $f_{\{0,6\}}$ are basically unchanged for all
pulse durations, the associated fragmentation process is ascribed to
the core-hole lifetimes which are independent of the pulse duration.

\begin{figure}
  \begin{center}
    \includegraphics[clip,width=\hsize]{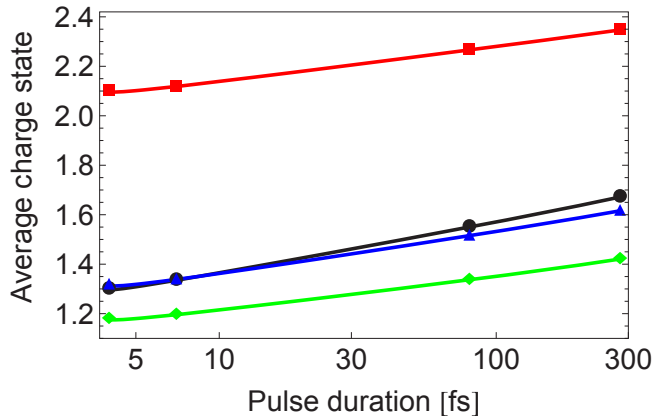}
    \caption{(Color online) Average charge state~$\tilde{\bar q}$ from~N$_2$
             subject to LCLS \xray~pulses of varying FWHM duration.
             We show~$\tilde{\bar q}$ deduced from the experimental ion
             yields of~N$_2$ [Fig.~\ref{fig:ionyields}]
             for the nominal pulse durations~$4$, $7$,
             $80$, and $280 \U{fs}$, alongside our calculations;
             the points are connected by interpolation.
             Specifically, we plot the experimental~$\tilde{\bar q}$
             as \textbf{\textcolor{mablack}{black}} circles, $\bar q$~from a
             single N~atom as \textbf{\textcolor{mared}{red}}
             squares [Sec.~\ref{sec:yiatom}], $\tilde{\bar q}$ from the
             symmetric-sharing model as \textbf{\textcolor{magreen}{green}}
             diamonds [Sec.~\ref{sec:lowerlim}], and
             $\tilde{\bar q}$ from the fragmentation-matrix model
             [Sec.~\ref{sec:fragmat}] as \textbf{\textcolor{mablue}{blue}}
             triangles.
             See Table~\ref{tab:fragconsts} for further LCLS~pulse parameters.}
    \label{fig:avcharge}
  \end{center}
\end{figure}

\begin{figure*}
  \begin{center}
    (a)~\includegraphics[clip,width=7cm]{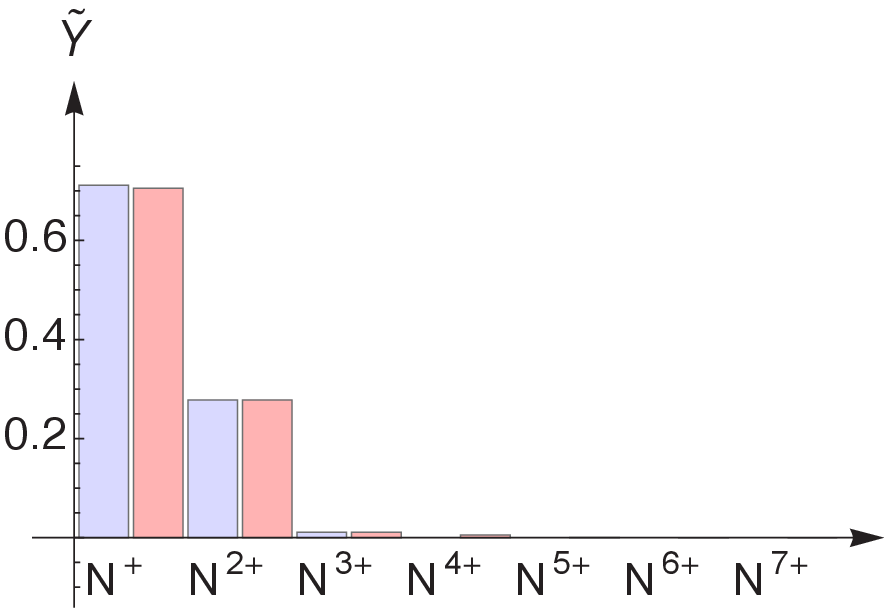}\hfill
    (b)~\includegraphics[clip,width=7cm]{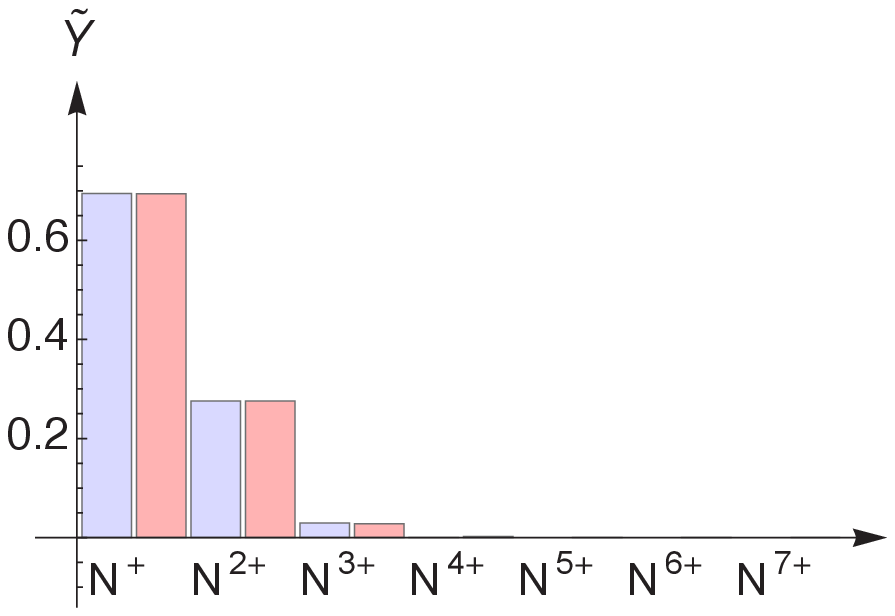}\\[8mm]
    (c)~\includegraphics[clip,width=7cm]{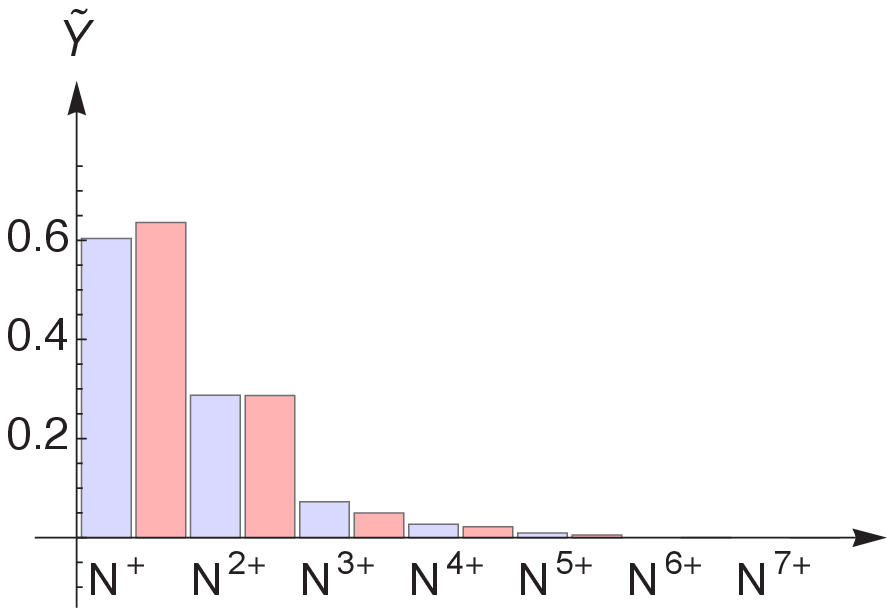}\hfill
    (d)~\includegraphics[clip,width=7cm]{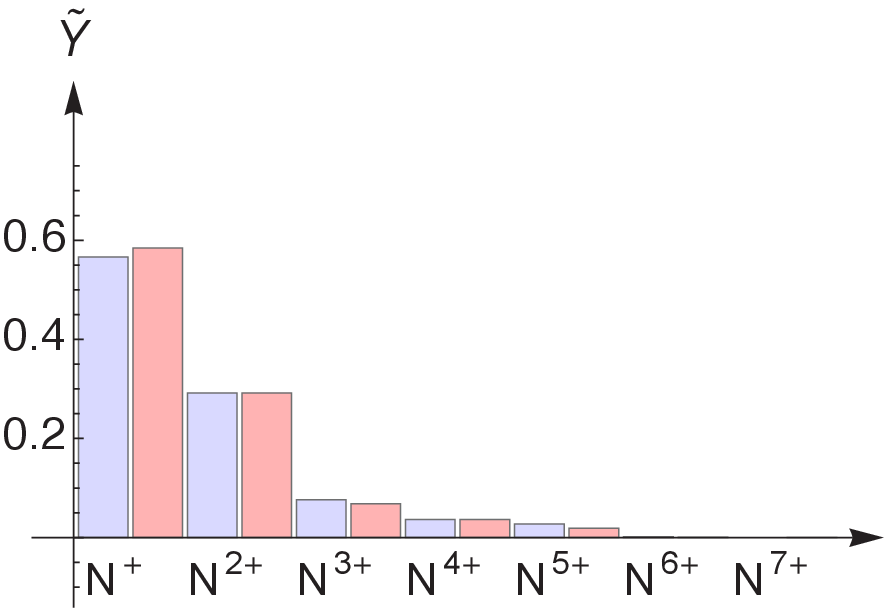}
    \caption{(Color) Ion yields~$\tilde Y_j$ of~N$_2$
             subject to LCLS \xray~pulses of varying duration:
             (a)~$4 \U{fs}$, (b)~$7 \U{fs}$, (c)~$80 \U{fs}$, and
             (d)~$280 \U{fs}$.
             Experimental~$\tilde Y_j$ are given by the
             \textbf{\textcolor{mablue}{blue}} bars and
             theoretical~$\tilde Y_j$ from the fragmentation-matrix model
             [Sec.~\ref{sec:fragmat}] by the \textbf{\textcolor{mared}{red}}
             bars.
             See Table~\ref{tab:fragconsts} for further LCLS~pulse parameters.}
    \label{fig:ionyields}
  \end{center}
\end{figure*}

The merits of the fragmentation matrix model are exhibited
in Figs.~\ref{fig:avcharge} and \ref{fig:ionyields} where we
plot~$\tilde{\bar q}$ and the ion yields, respectively, from the model
alongside the experimental data.
The agreement of our theoretical~$\tilde{\bar q}$ with the
experimental~$\tilde{\bar q}$
is very good, however, despite fitting the fragmentation matrix at all points,
the two~$\tilde{\bar q}$ do not coincide.
This has several reasons.
First, the fit is not arbitrary but physically motivated by the
molecular breakup process relying on the probabilities to find the atom
in various charge states.
The atomic rate equations of Sec.~\ref{sec:yiatom} were designed carefully,
yet only a one-electron approximation has been used and particularly shake off
is not treated.~\cite{Young:FE-10,Doumy:NA-11}
Other works on neon atoms~\cite{Young:FE-10,Doumy:NA-11}
show a far less than perfect agreement of atomic ion yields
from experiment with theory.
In the light of Refs.~\onlinecite{Young:FE-10,Doumy:NA-11},
our theoretical ion yields in Fig.~\ref{fig:ionyields}
appear almost too close to the experimental ones.

The finding that $\tilde{\bar q}$~from the fragmentation-matrix model
is lower than the experimental~$\tilde{\bar q}$ for longer pulses
is ascribed to the fact that no fragmentation during the pulse
is treated.
Namely, the fragmentation time of~$\mathrm N_2^{2+}$
is~\cite{Beylerian:CE-04}~${\sim}100 \U{fs}$
and thus shorter than the longest, $280 \U{fs}$, LCLS pulses.
Hence molecular fragmentation of~$\mathrm N_2^{2+}$ may
occur before the \xray~pulse is over.
When $\mathrm N_2^{2+}$ fragments into~$\mathrm N^+ + \mathrm N^+$,
absorption of another x~ray by the atomic ions~N$^+$ leads to an
enhancement of the $\mathrm N^{3+}$~yield;
fragmentation into~$\mathrm N + \mathrm N^{2+}$ increases the
$\mathrm N^{2+}$ and $\mathrm N^{4+}$~yields.
The fragmentation-matrix model does not distinguish this case
from breakup of the respective molecular ions which cannot be treated
by static ratios.
A time- and molecular-state-dependent fragmentation
matrix should be able to fully account for the observed
molecular ion yields.

Finally, we used the nominal pulse duration in all calculations so far
as was done in Ref.~\onlinecite{Hoener:FA-10}.
However, the pulse duration was found to be significantly shorter
in experiments.~\cite{Young:FE-10,Dusterer:FS-11}
Namely, a nominal pulse duration of~$80 \U{fs}$ was
estimated actually to be~$20$--$40 \U{fs}$ in Ref.~\onlinecite{Young:FE-10}
and measured to be~$40 \U{fs}$ in Ref.~\onlinecite{Dusterer:FS-11};
the nominal pulse duration of~$300 \U{fs}$~pulses was
found to be~$120 \U{fs}$.~\cite{Dusterer:FS-11}
Yet the experimental conditions in the two studies
Refs.~\onlinecite{Young:FE-10,Dusterer:FS-11} differ somewhat
from the one in Ref.~\onlinecite{Hoener:FA-10}.
Fortunately, the dependence of our theoretical predictions
on the pulse duration turns out to be quite low.~\cite{Buth:SU-up}

\section{Explanation of the observed phenomena in terms of elemental
molecular processes}
\label{sec:micromodel}

In the previous Secs.~\ref{sec:yiatom}, \ref{sec:lowerlim},
and \ref{sec:fragmat}, we investigated phenomenological models
that are based on the solution of the atomic rate equations
to describe the interaction of~N$_2$ with x~rays where
we used the average charge state~$\tilde{\bar q}$ [Eq.~(\ref{eq:avcharge})]
as a figure of merit to assess the quality of our models.
Here we discuss the problem from the perspective of the processes
entering the atomic rate equations to unravel the elemental molecular
processes that give rise to the behavior of the experimental~$\tilde{\bar q}$
in Fig.~\ref{fig:avcharge}.
In Refs.~\onlinecite{Hoener:FA-10,footnote1}, a molecular-rate-equation model
has been discussed based on elemental molecular
processes which differs in various aspects from the following
analysis.

\begin{figure*}
  \begin{center}
    \hfill(a)\includegraphics[clip,width=7cm]{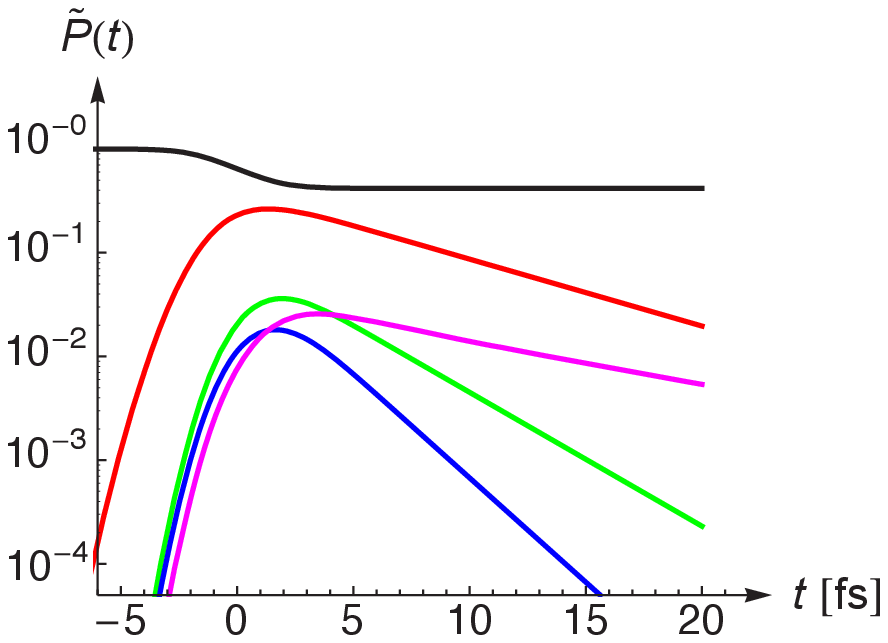}\hfill\hfill
          (b)\includegraphics[clip,width=7cm]{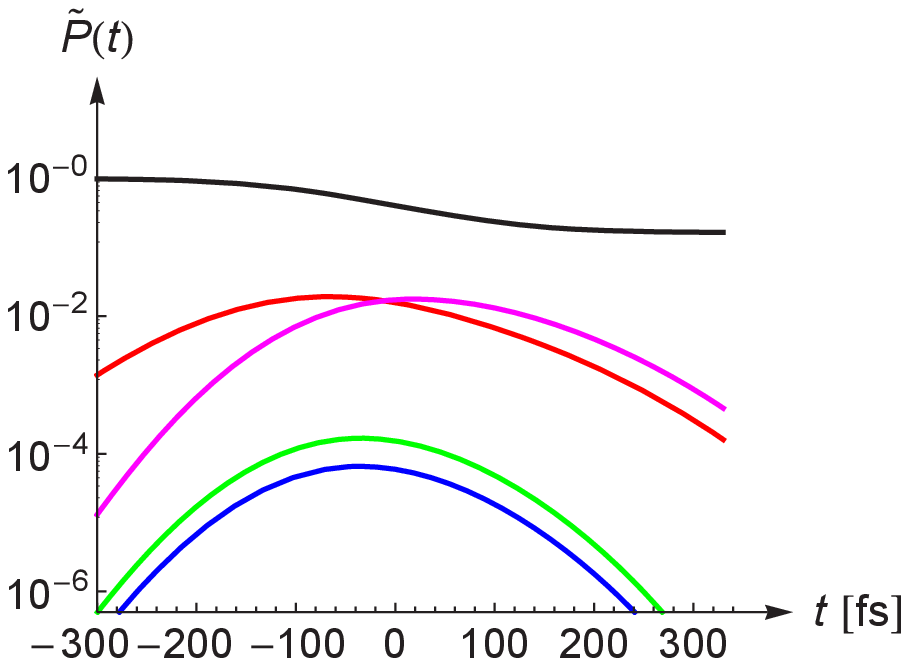}\hfill\
    \caption{(Color) The probabilities~$\tilde P(t)$ to
             find~N$_2$ during the \xray~pulse in its ground state
             is given by the \textbf{\textcolor{mablack}{black}} lines;
             the \textbf{\textcolor{mared}{red}} lines are the probability
             for a SCH;
             the \textbf{\textcolor{magreen}{green}} and
             \textbf{\textcolor{mablue}{blue}} lines stand for the
             probability to find a tsDCH and a ssDCH, respectively;
             the PAPs are indicated by \textbf{\textcolor{mamagenta}{magenta}}
             lines.
             Probabilities are determined for
             (a)~a Gaussian short pulse~(\ref{eq:XrayGaussEnv}) of a
             FWHM duration of~$4 \U{fs}$ and
             (b)~a Gaussian long pulse of a FWHM duration of~$280 \U{fs}$.
             See Table~\ref{tab:fragconsts} for further LCLS~pulse parameters.}
    \label{fig:domchannel}
  \end{center}
\end{figure*}

To describe the elemental molecular processes of~N$_2$ in intense x~rays,
we need to find out, first, which molecular electronic configurations prevail
in the course of the interaction with the \xray~pulse.
As the atomic ionization cross sections for core electrons are orders
of magnitude larger than cross sections for
valence electrons, predominantly core electrons are ionized.
The probabilities to find N$_2$ in its ground state or with a SCH,
a tsDCH, or a ssDCH can be deduced from the solution of the
atomic rate equations for an N~atom via~$P_{223}(t)^2$,
$2 \, P_{123}(t) \, P_{223}(t)$, $P_{123}(t) \, P_{123}(t)$,
and $P_{023}(t) \, P_{223}(t)$, respectively.
Furthermore, a sequence of photoionization~(P) with a following Auger
decay~(A) and another photoionization~(PAP) is a competing process
to DCHs for two-\xray-photon absorption.~\cite{Rohringer:XR-07,Buth:SU-up}
These probabilities are shown in Fig.~\ref{fig:domchannel} for a
short-pulse ($4 \U{fs}$) and a long-pulse ($280 \U{fs}$) scenario.
It is revealed clearly that for the long pulse [Fig.~\ref{fig:domchannel}b],
we have a contribution from almost exclusively SCHs and PAPs
with an almost vanishing probability to find DCHs.
On the contrary, for the short pulse, PAPs are less important
due to the fact that the Auger
decay time of a SCH~\cite{Buth:SU-up} of~$6.7 \U{fs}$ is longer than
the pulse duration.
Therefore, DCHs are more relevant for the short pulse
[Fig.~\ref{fig:domchannel}a] where tsDCHs make a larger contribution
than ssDCHs.
This is due to the bigger cross sections for the formation of tsDCHs compared
with ssDCHs and the longer lifetime of the former with respect to the latter.
Namely, for an N~atom in \xray~light at~$1100 \eV$ photon energy, the
cross section for SCH formation
%
%
is~$56 \U{kbarn}$ and it
%
%
is~$34 \U{kbarn}$ for DCH formation.
The SCH formation cross section is, therefore, by a factor
%
%
of~$1.7$ higher than the cross section for DCH~formation in an N~atom but
ssDCHs in~N$_2$, nevertheless, make a noticeable contribution for
short pulses, \ie, higher \xray~intensities, as is revealed
in Fig.~\ref{fig:domchannel}a.
The lifetime of a tsDCH is half the lifetime of a SCH,
%
%
\ie, $6.7 \U{fs} / 2 = 3.4 \U{fs}$,
%
%
versus~$2.2 \U{fs}$ for a ssDCH, respectively
[Tables~\tabAugerSingle{} and \tabAugerDouble{} of
Ref.~\onlinecite{Buth:SU-up}].
The conclusions drawn from Fig.~\ref{fig:domchannel} are
corroborated by the ratios in Table~\ref{tab:chprobs}
to find SCH~decay, first decay of DCHs, and PAPAs after
volume integration~(\ref{eq:beamprofile}).
We also show the impact of the SASE pulse
structure~\cite{Kondratenko:GC-79,Bonifacio:CI-84,Saldin:PF-00} on the ratios
using the partial coherence method to generate SASE
pulses~\cite{Pfeifer:PC-10,Jiang:TC-10,Cavaletto:RF-12}
with~$8 \eV$~bandwidth and a Gaussian temporal envelope that has the same FWHM
after squaring as the Gaussian pulse used to compute the reference ratios and
make Fig.~\ref{fig:domchannel}.
Inspecting Table~\ref{tab:chprobs}, we find that the impact of the SASE
pulse structure on the resulting ratios is small as has also been
found in previous work.~\cite{Rohringer:XR-07}

\begin{table*}
  \centering
  \begin{ruledtabular}
    \mbox{\begin{tabular}{cc|cccccc}
       $\tau\X{X}$ [$\mathrm{fs}$] & Pulse & SCH decay & ssDCH decay
                                   & tsDCH decay & PAPA \\
    \hline
       \phantom{00}4 & \phantom{00}1 SASE & 0.87 & 0.049 & 0.079 & 0.052 \\
                     & \phantom{0}10 SASE & 0.88 & 0.046 & 0.073 & 0.045 \\
                     &           100 SASE & 0.87 & 0.050 & 0.080 & 0.045 \\
                     &          1000 SASE & 0.87 & 0.050 & 0.081 & 0.046 \\
                     &           Gaussian & 0.89 & 0.041 & 0.065 & 0.035 \\
    \hline
                 280 & \phantom{00}1 SASE & 0.99 & 0.0035 & 0.0058 & 0.28 \\
                     & \phantom{0}10 SASE & 0.99 & 0.0045 & 0.0074 & 0.33 \\
                     &           100 SASE & 0.99 & 0.0044 & 0.0072 & 0.33 \\
                     &           Gaussian & 0.99 & 0.0043 & 0.0072 & 0.33
    \end{tabular}}
  \end{ruledtabular}
  \caption{Ratios to find SCH decay, first tsDCHs decay, the first ssDCHs decay,
           and a PAPA~process after averaging over a varying numbers of
           SASE pulses and from a Gaussian
           pulse with a FWHM duration of~$\tau\X{X}$.
           The ratios are probabilities from volume
           averaging~(\ref{eq:beamprofile})
           which are renormalized to the sum of the probabilities
           for (the first) decay of SCHs, tsDCHs, and ssDCHs.
           See Table~\ref{tab:fragconsts} for further LCLS~pulse parameters.}
  \label{tab:chprobs}
\end{table*}

Another reason why SCH, ssDCH, tsDCH, and PAP~channels have a particular
importance is the fact that N$_2^+$ and N$_2^{2+}$ are metastable
with respect to dissociation,~\cite{Wetmore:TO-86,Beylerian:CE-04}
\ie, the molecular ion remains essentially intact until it
has charged up to~N$_2^{3+}$ and higher.~\cite{Bandrauk:ES-99}
Therefore, these channels dominate the first stages of the interaction
of~N$_2$ with x~rays and thus have a crucial impact on fragmentation products
and such on molecular ion yields.
Similarly, for very high intensities, also N$_2$~with triple and
quadruple core holes may occur.
However, for our parameters, Fig.~\ref{fig:domchannel} reveals that we
need to focus only on a very limit number of initial molecular channels.
Based on the discussion of Secs.~\ref{sec:yiatom}, \ref{sec:lowerlim},
and \ref{sec:fragmat},
we deduce that one needs to consider following elemental molecular processes:

First, for long pulses with moderate \xray~photon flux, where the
probability to absorb two x~rays in a few-femtosecond time interval is low,
N$_2^{2+}$~frequently fragments into two atomic ions prior to
absorbing further x~rays.
However, for a higher \xray~flux this probability is not low and
differences may arise due to further absorption of x~rays prior to
molecular dissociation.
This introduces a time scale of~${\sim} 100 \U{fs}$
that determines the rate of fragmentation of the
dication~N$_2^{2+}$ [Ref.~\onlinecite{Beylerian:CE-04}].

Second, for high enough \xray~intensities,
further photoionization and Auger decay of the molecular
dication~N$_2^{2+}$ prior fragmentation forms molecular trications~N$_2^{3+}$.
No bound states were found for N$_2^{3+}$~ions in the midst
of many Coulomb repulsive potential
surfaces.~\cite{Safvan:DH-94,Bandrauk:ES-99}
Therefore, one may assume that those N$_2^{3+}$ without core holes
fragment into~N$^{2+}$ and N$^+$ which is the channel that is
indicated most prominently by the potential energy surfaces.
There is only a small amount of~N$_2^{3+}$ produced without core holes
in the present case due to small valence ionization cross sections for x~rays.
Although N$_2^{3+}$~is not metastable, dissociation is not immediate
and further x~rays may be absorbed prior dissociation.
Specifically, N$_2^{3+}$ with core holes should still be considered as a
molecular entity which is only broken up after Auger decay
because of the faster time scale of Auger decay compared
with molecular fragmentation.

Third, the sequence of two absorptions of an x~rays with subsequent
Auger decays~(PAPAs) or, likewise, the two Auger decays of a DCH
leads to~N$_2^{4+}$.
There is hardly any difference between the two situations
concerning the way molecular fragmentation takes place.
A SCH decay leads to~N$_2^{2+}$ which is metastable with a lifetime
that is large compared with inner-shell hole
lifetimes.~\cite{Wetmore:TO-86,Beylerian:CE-04,Buth:SU-up}
The absorption of a second x~ray then leads to~$\mathrm N_2^{3+}$
which is \emph{not} metastable~\cite{Safvan:DH-94,Bandrauk:ES-99}
and the molecular ion starts to dissociate, however, on a much
slower time scale than Auger decay.
The Auger decay time scale is independent of the \xray~pulse duration.
As shown in Table~\ref{tab:chprobs}, PAPAs are comparable with the first
decay of ssDCHs in the $4 \U{fs}$~case and PAPAs thus are an important
channel for two-\xray-photon absorption.
Furthermore, only a DCH is metastable with respect to dissociation;
as soon as the first Auger decay of a DCH occurs, $\mathrm N_2^{3+}$ is
formed which starts to dissociate in straight analogy to PAP
after the absorption of the second x~ray.
The second Auger decay of a DCH is on the same time scale as for the
PAPA process.
As both processes describe very similar situations with
respect to molecular fragmentation, they have the same
fragmentation pattern.
From our findings with the symmetric-sharing model
[Sec.~\ref{sec:lowerlim}] and results from laser-dissociation
experiments,~\cite{Codling:CS-91,Baldit:CE-05} we may assume
equal sharing of the four charges between the two nitrogen
nuclei for the fragmentation of N$_2^{4+}$~ions.
This is corroborated by inspecting Fig.~\ref{fig:ionyields}a
where no significant ion yield for~N$^{3+}$ is found which
would arise from a nonsymmetric sharing of charges
for the fragmentation of N$_2^{4+}$~ions resulting from DCHs or PAPAs.
In our case, DCHs and PAPs affect results for the short and long pulses
where the importance of DCHs is larger for the high \xray~flux of short pulses
because the molecules are ionized faster.
Fragmentation predominantly occurs after the short pulses are over.
For long pulses, the \xray~flux is reduced
leading to a slower rate of photoabsorption and frequently
fragmentation of~N$_2^{2+}$ occurs before further x~rays are absorbed.

Fourth, for even higher charge states,
dissociation becomes more and more rapid but the involved time scale
needs, nonetheless, be considered.
Namely, for \xray~pulses of only a few femtoseconds,
dissociation does not play a substantial role
and thus the assumptions underlying the fragmentation-matrix model
[Sec.~\ref{sec:fragmat}] are well fulfilled.
In this case, also for highly-charged~N$_2$, symmetric sharing should
be the predominant mode of fragmentation.~\cite{Safvan:DH-94}
For long \xray~pulses, breakup frequently occurs before high
molecular charge states are reached and subsequent \xray~absorption
occurs by independent ions.

The above analysis suggests that molecular effects and
molecular fragmentation of~N$_2$ in intense x~rays are centered
around these four steps.
The described elemental molecular processes should characterize the
most important aspects of the phenomenology of the interaction as
all insights gained from the analysis
of the models of Secs.~\ref{sec:yiatom}, \ref{sec:lowerlim},
and \ref{sec:fragmat} are incorporated.

\section{Conclusion}
\label{sec:conclusion}

In this paper, we discuss theoretically the interaction of intense,
short-pulse x~rays from the novel \xray~free electron laser Linac
Coherent Light Source~(LCLS) with N$_2$~molecules.
We set out from the computation of the energies, decay
widths, and one-\xray-photon absorption cross sections of a single
N~atom.
These data are input to describe the interaction an N~atom with
\xray~pulses in terms of a rate-equation model.
Ion yields and the average charge state are computed
and compared with experimental data for~N$_2$.
The agreement of theoretical data for an N~atom and experimental data for~N$_2$
is not satisfactory which we ascribe to a fragmentation of~N$_2$
after ionization causing a redistribution of charges over both atoms
in the molecule.
To describe the molecular degrees of freedom, we devise two
phenomenological models.
First, we compute a lower limit to the average charge state
assuming always equal sharing of charges between both atoms in~N$_2$.
Second, we use a fragmentation matrix to treat molecular dissociation
with static ratios to relate atomic and molecular ion yields.
The models clearly indicates the relevance of the redistribution of
charge and a different weighting of processes on short and long
time scales for similar nominal \xray~pulse energy.
The formation of high charge states at short pulse durations is suppressed
by a redistribution of charge in the multiply-ionized molecule
giving rise to the observed frustrated
absorption.~\cite{Hoener:FA-10}

Pertinent to our atomic rate equations of Sec.~\ref{sec:yiatom}, however,
is---as was shown by Young~\etal~\cite{Young:FE-10} for neon
atoms---that the deviation of the theoretical ion yields from
the experimental ones in the intense x~rays of LCLS
becomes more and more pronounced when the photon energy is
of the order of~$2000 \eV$.
The reason of this deviation is the fact that in Ref.~\onlinecite{Young:FE-10},
the rate equations for a neon atom contain only
all possible one-photon absorption and fluorescence and Auger decay channels
as is also the case for our rate-equation model of an N~atom
[Sec.~\ref{sec:yiatom}].
However, at such high photon energies, other channels involving many-electron
effects like shake off (for neon atoms in Ref.~\onlinecite{Doumy:NA-11})
may become important.
As in the present study of~N$_2$, the photon energy of~$1100 \eV$
also is high above the SCH-ionization threshold of an N~atom,
a similar deviation, as was found for neon,~\cite{Young:FE-10,Doumy:NA-11}
may be relevant.
However, we do not investigate this issue due to a lack of experimental data
on single N~atoms in intense x~rays.

Although our theoretical analysis is centered on~N$_2$, we believe
that many of our conclusions are generally true.
The time scale of the involved processes is dictated by the
formation and decay of SCHs and DCHs.
Higher numbers of core holes than one DCH per~N$_2$ are of little
importance at the present \xray~intensities.
Molecular fragmentation depends largely on the pacing of \xray~absorption
and, related, the abundance of SCHs versus DCHs.
For \xray~irradiation durations of only a few femtoseconds,
nuclear dynamics---specifically due to Coulomb explosion of the
charged fragments---does not play a role during the
\xray~pulse and we may assume fixed nuclei in the equilibrium geometry of
the molecule.
However, the nuclear wavepacket generated leads to a subsequent fragmentation
of the molecule.
Instead, for longer pulse durations, nuclear expansion during the \xray~pulse
becomes a crucial issue.
A molecule with a SCH may be metastable with respect to dissociation
and may even remain metastable after Auger decay has taken place.
Clearly, this is crucial for the understanding of the observed ion yields.
Hence molecular effects and molecular fragmentation are centered
around these initial steps.
For a general description of molecules in intense x~rays, one needs to
account for the complicated
joint problem of \xray~absorption, decay processes, and nuclear dynamics.
We make first attempts towards a full treatment of the problem.

Based on our studies, many fascinating possibilities for future research
open up.
An investigation with \emph{ab initio} methods of molecular
physics~\cite{Szabo:MQC-89} of the absorption of x~rays, the
decay of inner-shell holes and the resulting nuclear dynamics
poses a challenge.
Nonetheless, we consider it of high importance to find an
\emph{ab initio} description to gain deeper insights
and to establish a theoretical foundation for molecules
in intense x~rays.
The formation of DCHs gives one detailed information about the structure
of a molecule;
it was found to be more sensitive than
SCH~spectroscopy.~\cite{Cederbaum:DV-86,Santra:XT-09,Tashiro:MD-10,Berrah:DC-11}
The time-delay between the two core-hole ionization
events in DCH~production needs to be short enough, such that the first
vacancy has not decayed.
Then the second photoelectron carries detailed information
about the dicationic states of the molecule.
This gives rise to \xray~two-photon photoelectron
spectroscopy~(XTTPS).~\cite{Santra:XT-09}
Furthermore, we have a valence electron density in molecules that may
differ appreciably from the density produced by the noninteracting
atoms in the molecular geometry.
This may in some cases have a noticeable influence on the
rate of intraatomic electronic decay~(IAED),~\cite{Buth:IM-03}
if valence electrons are involved.
Additionally, the close packing of the atoms in a molecule gives rise
to interatomic electronic decay processes for inner-valence holes,
namely, interatomic Coulombic decay~(ICD) and electron transfer mediated
decay~(ETMD) (Ref.~\onlinecite{Buth:IM-03} and References therein).
The relevance of these additional channels needs to be assessed
in future work.

\begin{acknowledgments}
C.B.~would like to thank Nikolai V.~Kryzhevoi for fruitful discussions
and a critical reading of the manuscript.
We are grateful to Oleg Kornilov and Oliver Gessner for
discussions on nitrogen molecules in intense x~rays.
C.B.~was supported by the National Science Foundation under
grant~Nos.~PHY-0701372 and PHY-0449235 and by a Marie Curie International
Reintegration Grant within the 7$^{\mathrm{th}}$~European Community
Framework Program (call identifier: FP7-PEOPLE-2010-RG, proposal No.~266551).
J.-C.L.~thanks for support by the Fundamental Research Funds for the
Central Universities.
Additional funding was provided by the Office of Basic Energy Sciences,
Office of Science, U.S.~Department of Energy,
for~C.B.~under Contract No.~DE-AC02-06CH11357 and
for~L.F., M.H., and N.B.~under Contract No.~DE-FG02-92ER14299.
M.H.C.'s~work was performed under the auspices of the U.S.~Department
of Energy by Lawrence Livermore National Laboratory under
Contract No.~DE-AC52-07NA27344.
R.N.C.~is supported through the LCLS at the SLAC National Accelerator
Laboratory by the U.S.~Department of Energy, Office of Basic Energy Sciences.
J.P.C.~and J.M.G.~are supported through both the LCLS and
The PULSE Institute for Ultrafast Energy Science at the
SLAC National Accelerator Laboratory by the U.S.~Department of Energy,
Office of Basic Energy Sciences.
Portions of this research were carried out at the Linac Coherent Light
Source~(LCLS) at SLAC National Accelerator Laboratory.
LCLS is an Office of Science User Facility operated for the
U.S.~Department of Energy Office of Science by Stanford University.
\end{acknowledgments}


\begin{thebibliography}{67}
\bibitem [{foo()}]{footnote1}%
  \BibitemOpen
  \href@noop {} {}\bibinfo {note} {The phenomenological model for~N$_2$
  discussed in Ref.~\onlinecite{Hoener:FA-10} is based on molecular rate
  equations and was designed by Oleg Kornilov and Oliver Gessner.}\BibitemShut
  {Stop}%
\end{thebibliography}
\end{document}